\documentclass[11pt,a4paper]{article}

\pdfoutput=1

\usepackage{jheppub}

\usepackage{subfigure}

\usepackage{feynmp}
\usepackage{hyperref,graphicx}
\usepackage{xcolor,pict2e,tikz}
\usetikzlibrary{patterns,arrows}

\def\gsim{\mbox{~{\raisebox{0.4ex}{$>$}}\hspace{-1.1em}
  {\raisebox{-0.6ex}{$\sim$}}~}}

\def\ie{\textsl{i.e.}}
\def\Eq#1{Eq.~(\ref{#1})}
\def\Dim{\text{D}}
\def\Nc{N}
\def\SUN{\text{SU}(\Nc)}

\def\bigO{\mathcal{O}}
\def\ep{\epsilon}

\def\YM{\text{YM}}
\def\HIG{\phi}

\def\Tr{\text{Tr}}

\def\be{\begin{equation}}
\def\ee{\end{equation}}
\def\bea{\begin{eqnarray}}
\def\eea{\end{eqnarray}}
\def\MSbar{\overline{\mbox{MS}}}
\def\mubar{\bar{\mu}^2}
\def\Tproj{{\mathbf T}}
\def\Lproj{{\mathbf L}}

\def\GT{G_T}
\def\GL{G_L}
\def\Dhig{D_{-}}
\def\Dsym{D_{+}}
\def\Dcond{\mathcal{D}}
\def\bphi{\bar\phi}
\def\ansatz{\textit{Ansatz}}
\def\ansatze{\textit{Ans\"atze}}
\def\ep{\epsilon}

\def\IR{\text{IR}}
\def\UV{\text{UV}}

\def\fmfcon{\fmfv{decor.shape=circle,decor.filled=full,decor.size=0mm}}
\def\fmfIR{\fmfv{decor.shape=circle,d.filled=10,d.size=5mm,l=$\Pi$,l.dist=0mm}}

\title{$n$PI Resummation in 3D SU($N$) Higgs Theory}
\author{Mark C. Abraao York,}
\author{Guy D. Moore}
\affiliation{McGill University Department of Physics\\
3600 Rue University\\
Montr\'eal, QC\\
H3A 2T8}

\abstract{
We test the utility of the $n$PI formalism for solving nonperturbative
dynamics of gauge theories by applying it to study the phase diagram of
$\SUN$ Higgs theory in 3 Euclidean spacetime dimensions.  Solutions
reveal standard signatures of a first order phase transition with a
critical endpoint leading to a crossover regime, in qualitative agreement
with lattice studies.  The location of the
critical endpoint, $x \sim 0.14$ for SU(2) with a fundamental Higgs,
is in rough but not tight quantitative agreement with the lattice.
We end by commenting on the overall effectiveness and limitations of an $n$PI
effective action based study.  In particular, we have been unable to
find an $n$PI gauge-fixing procedure which can simultaneously display
the right phase structure and correctly handle the large-VEV Higgs region.
We explain why doing so appears to be a serious challenge.}

\begin{document}

\maketitle

\section{Introduction}

Thermal or off-equilibrium dynamics of hot nonabelian gauge theories have
applications in heavy ion physics (see e.g. \cite{Adams:2005dq} and references therein) and in early Universe
cosmology \cite{Cohen1,Rubakov1}.  An important feature of nonabelian
gauge theory is asymptotic freedom, which makes the coupling smaller at
shorter length (or higher energy) scales.  Naively this means that
perturbation theory should work better at higher temperatures, where the
relevant energy scale $T$ is large.  In fact this is only partly the
case.  As shown already in 1980 \cite{Linde1,Gross:1980br}, the
behavior on scales $\ell > 1/T$ is in fact that of a 3-dimensional
theory, which goes rapidly to strong coupling at longer distances.
Therefore the long-distance $\ell \gg T$ behavior of nonabelian gauge
theory is strongly coupled at \textsl{any} temperature.

There is a dearth of tools for computing real-time dynamics of
nonabelian gauge theories when nonperturbative physics is involved.
This frustrates efforts to understand dynamics, both at strong coupling
and at weak coupling.  A method which has shown much promise in scalar
and Yukawa theories is the $n$-particle irreducible ($n$PI) method
\cite{Baym,CJT,Reinosa,Aarts1,Braaten3PI,Berges2PI,Berges2}.  While the motivation for developing such methods
lies largely in the hopes that they can be applied to nonabelian gauge
theories \cite{Berges1,Blaizot1}, almost no work in that direction has occurred yet.  There have
been some results in abelian theories \cite{Reinosa1,Borsanyi1,Carrington2,Andersen2},
and some arguments as well as a computation demonstrating
that a 3-particle-irreducible treatment of QCD would automatically
capture the leading perturbative effects relevant for transport and
equilibriation \cite{Carrington}.  But no one has made a concerted effort
to apply the $n$PI approach to nonabelian gauge dynamics.

In a previous paper \cite{3PI} we took a first step in this
direction, by applying the 3-particle irreducible (3PI) method to the
study of Yang-Mills theory in three Euclidean dimensions.  The main
motivation was to test out the methodology in a context where we
\textsl{do} have other computational tools at our disposal, so we can
appraise whether it is successful before undertaking the more
challenging problem of applying 3PI methods to dynamics.  But if
successful, the 3PI method could still have real utility as a
potentially faster or more efficient method of studying 3D theories,
which are in fact of intrinsic interest.  In particular, the 3D theory
captures the large-distance nonperturbative physics of weakly-coupled
hot gauge theories alluded to above, at least at the thermodynamical
level.

Unfortunately, in that work we were only able to solve the 3PI problem
for pure gauge theory in 3 dimensions.  There are few long-distance
sensitive observables in that theory, so we lacked gauge-invariant
measurements to match to (lattice) nonperturbative calculations in 3D
Yang-Mills theory and did not find effective ways
to test whether the method ``works.''  In the present paper we intend to
address this by extending our $n$PI treatment to 3-dimensional Yang-Mills
Higgs theory, a theory which has a nontrivial phase structure.  We will
study whether the $n$PI approach can successfully reproduce the phase
structure of the theory, a nontrivial and nonperturbative test of the
technique.  We emphasize that our purpose is as a test of whether the
$n$PI approach in gauge theory can reproduce nonperturbative phenomena.
The goal is not to improve our understanding of 3D Yang-Mills Higgs
theory, which has been thoroughly studied using lattice gauge theory
techniques \cite{Kajantie2,Kajantie3,endpoint}.
In the remainder of the introduction, we will explain both
3D Yang-Mills Higgs theory and the $n$PI approach in a little more
detail.

The study of three dimensional nonabelian gauge theory is motivated by
its relationship to electroweak theory and QCD via dimensional reduction
\cite{Appelquist1,Farakos1,Kajantie5,Braaten1,Kajantie1}. At high
temperature $T \gg \Lambda_{\text{QCD}}$, QCD exhibits a natural
separation of scales $g^2 T \ll g T \ll T$ so that non-zero bosonic and
all fermionic Matsubara modes become heavy compared to the soft scales
of the theory. These modes can be integrated out to obtain an effective
3 dimensional description of the soft physics, which is precisely SU(3)
Yang Mills coupled to an adjoint scalar with gauge coupling $g^2_{3\Dim}
= g^2_{4\Dim}T$ and mass $m^2_A = g^2(\Nc/3 + N_f/6)T^2$ (identified
with the 0-mode of the $A_0$ component of the 4D gauge field). If one is
only interested in physics at the supersoft scales, this can be taken
one step further by integrating out the $A_0$ field to yield pure 3D
Yang-Mills.

Yang-Mills theory, QCD and electroweak theory are known to undergo a
phase transition \cite{Pisarski1,Svetitsky1,Fradkin1,Kirzhnits1}
over certain ranges of the model parameters. Naturally, for physical
values of these parameters, one would ask whether we are in a first
order, second order or crossover regime. 3D effective models could
potentially shed some light on this matter, except that for QCD, where
the effective 3D description is an $\text{SU}(3) + \text{adjoint Higgs}$
theory, $T_c \sim \Lambda_{\text{QCD}}$. Thus, in the vicinity of the
QCD phase transition (or crossover) the effective description breaks
down, since the underlying assumption of weak 4D coupling and a
separation of scales is not valid. A 3D effective model is still useful
for studying the nonperturbative infrared dynamics of hot QCD, just not
at temperatures in the vicinity of the scale
$\Lambda_{\text{QCD}}$. However, the situation is different for
electroweak theory near its phase transition.

The effective 3D description of electroweak theory resulting from
dimensional reduction is an
$\text{SU}(2) \times \text{U}(1)$ gauge theory coupled to both
fundamental and adjoint scalars. The adjoint scalars arise via the
dimensional reduction, while the fundamental scalar is identified with
the 4D Higgs field. In practice, an accurate study of the 4D theory does
not require such elaborate field content; rather, quantitative
predictions can be made by considering the much simpler $\text{SU}(2) +~
\text{fundamental}$ case
\cite{endpoint,Kajantie2,Kajantie3,Gurtler1,Ritz1}. Then, as a
further refinement, one may study the effects due to the inclusion of an
adjoint field \cite{Farakos2,Kajantie6}, as well as a $\text{U}(1)$
gauge field \cite{Kajantie4}. Or, in the context of GUTs, the model with
an SU(5) or $\text{SU}(3) \times \text{SU}(2)$ gauge group may be of
interest \cite{Rajantie1,Laine1}. These models have received a fair
amount of attention in the past due to the significance of a phase
transition on electroweak baryogenesis \cite{Cohen1,Rubakov1}.
For the models considered, a first order phase transition at physical values of
the Higgs mass has been ruled out.

\begin{figure}
\centering
\begin{tikzpicture}[
    x=2.5mm,y=2.5mm,
    axis/.style={very thick, ->, >=stealth'},
    ]
    % axis
    \draw[axis] (-1,0)  -- (21,0) node[right] {\Large $x$};
    \draw[axis] (0,-1) -- (0,11) node[above] {\Large$xy$};
	\draw (12,5) node[right]{\Large$(x_c,y_c)$};
	\draw [line width=1,color=black] (0,5) -- (12,5);
	\draw [line width=1,color=black,dashed] (0,2) -- (12,5);
	\draw [line width=1,color=black,dashed] (0,8) -- (12,5);
	\draw (2,5) node[above]{$y_c(x)$};
	\draw (4,8) node[right]{$y_{+}(x)$};
	\draw (4,2) node[right]{$y_{-}(x)$};
\end{tikzpicture}
\caption{\label{fig:phasediagram}Phase diagram for SU($\Nc$) Higgs
  theory, showing a first order line terminating at a critical
  point. The dashed lines indicate the appearance of metastable
  configurations in the effective potential.}
\end{figure}
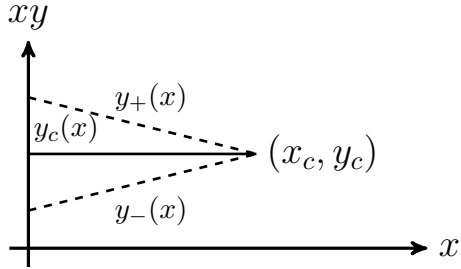

We show a cartoon of the phase diagram of $\SUN$ fundamental-Higgs theory in
Fig.~\ref{fig:phasediagram}.  It is parametrized by two dimensionless
variables $x$ and $y$, describing the scalar self-coupling and
renormalized mass respectively, normalized to the appropriate power of
$g^2$.  In terms of the 4D thermal theory, $x$ is predominantly set by
$m_{\text{H}}/m_{\text{W}}$ and $y$ is predominantly set by the
temperature.  Therefore an evolution in temperature in the 4D theory
corresponds to a nearly vertical line on the phase diagram; moving
horizontally is changing the vacuum parameters of the theory.
The diagram has a first order line which ends at a second-order 3D Ising
universality \cite{endpoint} endpoint at $(x_c,y_c)$; if $y$ is varied
holding $x<x_c$ fixed, the system encounters a first order phase
transition. But the transition is not a symmetry-breaking phase
transition, and no order parameter distinguishes one phase from the
other, similar to the liquid-gas phase transition (which is in the same
universality class). We have also indicated upper and lower
metastability lines, which show how deeply the system can ``superheat''
or ``supercool'' before encountering spinodal instability.  The
locations of these lines cannot be rigorously defined; but they will enter in our analysis so we indicate them for completeness.

At small $x$ the phase transition can be studied perturbatively by
computing the one-loop effective potential for the Higgs vacuum
expectation value (VEV). Indeed one
finds that in this region, the phase transition is first order. However,
the perturbative treatment then goes on to predict a first order phase
transition for all values of $x$! Higher-order computations
\cite{Arnold2,Hebecker} show that the perturbative expansion
parameter is actually $x$, indicating that perturbation theory fails for
large values of $x$, which turns out to mean $x \gsim 1/10$.

Therefore, the end point and crossover must be resolved
\textit{nonperturbatively}. Since the lattice has already provided us
with a very accurate determination of the phase diagram, we are able to
use these results to test the reliability and accuracy of an alternative
nonperturbative approach to the lattice, namely that of $n$PI
resummation in a gauge theory setting. In this
paper we will study the application of the $n$PI (specifically 2PI)
formalism to $\SUN$ Higgs theory.

In the context of a hot gauge theory, the use of an
$n$PI based resummation scheme is primarily motivated by the extremely
poor convergence of a weak-coupling expansion \cite{Andersen1}, since it
provides a systematic procedure for reorganizing a perturbation
series. Our approach here is along a trajectory which differs from many of the previous works on this subject mentioned earlier. That is, by applying the $n$PI
formalism to $\SUN$ Higgs theory our goal is to directly solve the
resulting self-consistent, Schwinger-Dyson (SD) resemblant\footnote{The integral equations of motion obtained via the $n$PI formalism are not strictly speaking Schwinger-Dyson equations. However, they are qualitatively similar, so we will often refer to them as SD equations throughout the text.} integral equations in a manner
reminiscent of \cite{3PI}, and then subsequently derive gauge-invariant
quantities from the solutions.

An $n$PI effective action $\Gamma[\bphi,G,...]$ generates equations of
motion for $n$-point resummed vertices
by variation with respect to these $n$-point functions. We will
specifically consider a three-loop truncation of the case $n = 2$, which
in terms of diagrams can be interpreted as resumming one- and two-loop
self-energy topologies. (In \cite{3PI} we treated the pure-gauge theory
at the $n=3$ level, that is, including as well a self-consistent one-loop
resummation of three-point vertices.  The result established that the
corrections to these vertices are small, so we have avoided this
technical complication.)  By solving the resulting ``SD'' equations, we
can subsequently compute the gauge-invariant scalar condensate $\langle
\phi^{\dagger}\phi\rangle$ as a function of the parameters $x$ and $y$
on the phase diagram. Then, at a specific value $x$, from the behavior
of $\langle \phi^{\dagger}\phi\rangle$ over a range of $y$ we can infer
whether or not we are in the crossover or first order phase transition
region. This will allow us to bracket and locate the critical end
point.

We will present the technical details of the computation for a single
complex scalar field in representation $R$ of $\SUN$. Results will be
given for $\Nc=2$ (fundamental representation) in Landau and Feynman
gauges, however it should be noted that the method straightforwardly
generalizes to the inclusion of additional scalar fields, higher
representations, and larger gauge groups.

The text is organized as follows: in Section \ref{SUNHiggs} we will
present the three-loop truncated 2PI effective action for $\SUN$ Higgs
theory, as well as the SD equations that it generates. Additionally,
some details pertinent to regularization and renormalization will be
reviewed here. In Section \ref{Extremization} we will present certain
extensions to the algorithm described in \cite{3PI} which are needed to
solve the 2PI equations of motion numerically. In Section \ref{Results}
we will give an overview of the results, as well as derived quantities
such as the scalar condensate and the location of the critical end
point. Finally, throughout Sections \ref{Extremization} and \ref{Results} we will also discuss the
properties of the effective action, and comment on the overall
effectiveness of the method.

\section{$\SUN$ Yang-Mills + Higgs theory in the $n$PI formalism}

\label{SUNHiggs}
\subsection{General remarks and notation}

It is useful to begin by reviewing a number of the basic conventions
that are used throughout. It should be assumed that $T_R^a$ is a
generator of some representation $R$ of $\SUN$. The fundamental and
adjoint representations are denoted by $F$ and $A$ respectively, and
$d_R$ is the dimension of $R$, for instance $d_F = C_A = \Nc$. We have
\bea
\Tr ~ T_R^aT_R^b &=& T_R \delta^{ab} \\
 T^s_{Rim} T^s_{Rmj} &=& C_R \delta_{ij} ,
\eea
and additional group theory identities needed in this computation can
be found in \cite{vanRitbergen}. Following gauge fixing, the
Lagrangian can be divided into a Yang-Mills component and a Higgs
component,
\bea
\label{eq:SYM}\mathcal{L}_{\text{YM}} &=& \frac{1}{2}\Tr
F_{\mu\nu}F^{\mu\nu} +\frac{1}{2\xi}(\partial^\mu A_\mu^a)^2 +
\partial_\mu\bar{c}^a \partial^\mu c^a - gf_{abc}\partial^\mu
\bar{c}^a c^b A_\mu^c\\
\mathcal{L}_{\HIG} &=& (D^\mu \phi)^\dagger (D_\mu\phi) + (m^2 +
\delta m^2) \phi^\dagger \phi +
\lambda(\phi^\dagger\phi)^2 \label{eq:SHiggs}
\eea
so that $\mathcal{L} = \mathcal{L}_{\text{YM}} + \mathcal{L}_{\HIG}$
(in general covariant gauge as written).  Note that we have not fixed
to $R_\xi$ gauge; our gauge fixing only acts on the gauge degrees of
freedom.  We will discuss this more in Subsection \ref{gaugefix}.

We define the dimensionless ratios
\be
x = \frac{\lambda}{g^2} \qquad y=\frac{m^2}{g^4} ,
\ee
which is the same as the definition introduced in Ref.~\cite{Kajantie2}
and commonly used throughout the literature. In \Eq{eq:SHiggs}
an additive counterterm has been explicitly included to cancel the
divergent two-loop self-energy graphs (its value is given in Appendix
\ref{SelfEnergy}). This leads to a scale dependence in $m^2$ and $y$
accordingly; for a fundamental SU(2) Higgs we have
\bea
\frac{d y}{d \log \mu} &=& -\frac{1}{16 \pi^2} \Big (\frac{51}{16} +
9x - 12 x^2 \Big ) .
\eea
The mass renormalization scale is fixed at $\mu = g^2$ throughout. For
phenomenological applications, the relation between $x,y$ and 4D
theory parameters are \cite{Kajantie2}
\bea
x &=& -0.00550 + 0.12622 \Big ( \frac{m_{\text{H}}}{80.6 \text{GeV}}
\Big )^2 \label{eq:x}\\
y &=& 0.39818 + 0.15545  \Big ( \frac{m_{\text{H}}}{80.6 \text{GeV}}
\Big )^2 - 0.00190  \Big ( \frac{m_{\text{H}}}{80.6 \text{GeV}} \Big
)^4 - 2.58088 \frac{m^2_{\text{H}}}{T^2} \label{eq:y} \quad
\eea
assuming a value of $g_{\text{4D}} = 2/3$ for the 4D gauge coupling.

\subsection{The three-loop 2PI effective action}
The 2PI effective effective action $\Gamma[G_{ij}]$ is formally
defined as the Legendre transform of the generating function of
connected diagrams $W[K_{ij}]$ with respect to a two particle source
\cite{CJT}. Using the generic label $\Phi_i$ for fields, $W[K_{ij}]$
reads
\be \label{eq:generatingfunction}
W[K_{ij}] = - \log \int D[\Phi] e^{-S - \frac{1}{2}\Phi_i K_{ij} \Phi_j} .
\ee
Even correlation functions can be obtained by differentiation with
respect to $K_{ij}$. For instance,
\be\label{eq:dGdK}
\frac{\delta W[K_{ij}]}{\delta K_{ij}} = \frac{1}{2} G_{ij}
\ee
yields the two-point function $G_{ij}$. For the two-point functions of
$\SUN$ Higgs theory, we can assume that $G_{ij}$ is proportional to
the color identity of the corresponding species, and hence so is
$K_{ij}$. Then, for a rotationally symmetric Lagrangian
\be
\langle \Phi_{i}\rangle =  \frac{\int D[\Phi] ~\Phi_i ~ e^{-S - \frac{1}{2}\Phi_i K_{ij} \Phi_j} }{\int D[\Phi] ~ e^{-S - \frac{1}{2}\Phi_i K_{ij} \Phi_j}} = 0 ,
\ee
\ie, the presence of $K_{ij}$ does not alter the global rotational
invariance of the original Lagrangian. So in fact, \Eq{eq:dGdK}
generates the \textit{connected} two-point function. The consequences
of this statement in the context of a spontaneously broken gauge
theory will be discussed towards the end of this section, but for now
we can proceed with the Legendre transform
\be
\Gamma[G_{ij}] = K_{ij}\frac{\delta W[K_{ij}]}{\delta K_{ij}} - W[K_{ij}] .
\ee
In setting $K_{ij} = 0$, equations of motion for $G_{ij}$ are obtained
from the stationarity condition
\be
\frac{\delta\Gamma[G_{ij}]}{\delta G_{ij}} = 0 .
\ee
The solutions we seek correspond to extrema of
$\Gamma[G_{ij}]$. Specializing now to the field content of $\SUN$
Higgs theory, we can write $\Gamma = \Gamma_{\YM} + \Gamma_{\HIG}$ and
explicitly state the loop expansion, which we will truncate at
three loops. $\Gamma_{\YM}$ is defined so that it contains those
diagrams encountered in the pure Yang-Mills problem while
$\Gamma_{\HIG}$ contains the additional diagrams which arise when a
single arbitrary representation Higgs field is included. Using the
diagrammatic notation
\begin{fmffile}{fmf_propagators}
\bea
G_{\mu\nu} &=& ~\parbox{20mm}{\begin{fmfgraph}(20,10)
\fmfleft{l1}\fmfright{r1}
\fmfforce{0.5w,0.5h}{v1}
\fmf{dbl_wiggly}{l1,r1}
\end{fmfgraph}}\\
\Delta &=&  ~\parbox{20mm}{\begin{fmfgraph}(20,10)
\fmfleft{l1}\fmfright{r1}
\fmfforce{0.5w,0.5h}{v1}
\fmf{dbl_dots_arrow}{l1,r1}
\end{fmfgraph}}\\
D &=& ~\parbox{20mm}{\begin{fmfgraph}(20,10)
\fmfleft{l1}\fmfright{r1}
\fmfforce{0.5w,0.5h}{v1}
\fmf{dbl_plain_arrow}{l1,r1}
\end{fmfgraph}} ~ ,
\eea
without loss of generality we write the two-point functions as
\bea
\label{eq:Dfull}G_{\mu\nu}(p) &=&
\frac{1}{p^2-\Pi_T(p)}\mathbf{T}_{\mu\nu} + \frac{\xi}{p^2 -
  \xi\Pi_L(p)} \mathbf{L}_{\mu\nu}\,, \\
\label{eq:Deltafull}\Delta(p) &=& \frac{1}{p^2 - \Sigma(p)}\,, \\
\label{eq:Gfull}D(p) &=& \frac{1}{p^2 + m^2 - \Pi_{\HIG}(p)} \,.
\eea
All vertices appearing in the 2PI effective action are at
tree level. These are drawn as
\vspace{5mm}
\be
\parbox{15mm}{\begin{fmfgraph*}(15,15)
\fmfleft{l1,l2}\fmfright{r1,r2}
\fmflabel{$p_3,a_3,\mu_3$}{l1}\fmflabel{$p_4,a_4,\mu_4$}{l2}
\fmflabel{$p_1,a_1$}{r1}\fmflabel{$p_2,a_2$}{r2}
\fmf{dbl_wiggly}{l1,v1}\fmf{dbl_wiggly}{l2,v1}
\fmf{dbl_plain}{r1,v1}\fmf{dbl_plain_arrow}{v1,r2}
\end{fmfgraph*}} \quad =
g^2\mathcal{V}^{a_1a_2a_3a_4}_{\mu_3\mu_4}(p_1,p_2,p_3,p_4)\vspace{5mm}
\ee
with the corresponding expressions given in Appendix
\ref{FeynRules}. Finally, the Higgs mass renormalizes at two loops; it
is necessary to subtract the divergence with an additive counterterm
of the form $m^2 = m_{\HIG}^2 + \delta m^2$, with the corresponding
vertex
\be
\parbox{20mm}{\begin{fmfgraph}(20,10)
\fmfleft{l1}\fmfright{r1}
\fmfforce{0.5w,0.5h}{v1}
\fmfv{decoration.shape=hexacross}{v1}
\fmf{dbl_plain}{l1,v1}
\fmf{dbl_plain}{v1,r1}
\end{fmfgraph}} = -\delta m^2 .
\ee\end{fmffile}
Explicitly factoring out minus signs due to ghost loops, we have
\begin{fmffile}{fmf_gammaYM}
\begin{eqnarray}
\label{eq:GammaYM}
\nonumber \Gamma_{\text{YM}} &=& \frac{1}{2} \Tr \log G
- \frac{1}{2} \Tr G^{(0)}_{\mu\nu} G^{\mu\nu}
- \Tr\log\Delta
+ \Tr [\Delta^{(0)}]^{-1}\Delta \\
\nonumber &&+~ \frac{1}{12}~
\parbox{20mm}{\begin{fmfgraph}(20,20)
\fmfforce{1mm,0.5h}{v1}\fmfforce{19mm,0.5h}{v2}
\fmf{dbl_wiggly,left,tension=0.4}{v1,v2,v1}
\fmf{dbl_wiggly}{v1,v2}
\fmfcon{v1}
\end{fmfgraph}} ~+ \frac{1}{8}~
\parbox{20mm}{\begin{fmfgraph}(20,20)
\fmfforce{1mm,0.5h}{v1}\fmfforce{19mm,0.5h}{v3}
\fmfforce{0.5w,0.5h}{v2}
\fmf{dbl_wiggly,right,tension=0.4}{v1,v2,v1}
\fmf{dbl_wiggly,left,tension=0.4}{v3,v2,v3}
\end{fmfgraph}}
~- \frac{1}{2}~
\parbox{20mm}{\begin{fmfgraph}(20,20)
\fmfforce{1mm,0.5h}{v1}\fmfforce{19mm,0.5h}{v2}
\fmf{dbl_dots_arrow,left,tension=0.4}{v1,v2,v1}
\fmf{dbl_wiggly}{v1,v2}
\fmfcon{v1}
\end{fmfgraph}}\\
\nonumber &&+~ \frac{1}{48}~
\parbox{20mm}{\begin{fmfgraph}(20,20)
\fmfforce{1mm,0.5h}{v1}\fmfforce{19mm,0.5h}{v2}
\fmf{dbl_wiggly,left,tension=0.4}{v1,v2,v1}
\fmf{dbl_wiggly,left=0.5}{v1,v2}
\fmf{dbl_wiggly,right=0.5}{v1,v2}
\end{fmfgraph}} ~+ \frac{1}{24}~
\parbox{20mm}{\begin{fmfgraph}(20,20)
\fmfforce{10mm,10mm}{v4}\fmfforce{2.206mm,5.5mm}{v1}
\fmfforce{17.794mm,5.5mm}{v2}
\fmfforce{10mm,19mm}{v3}
\fmfcon{v1,v2,v3,v4}
\fmf{dbl_wiggly,left=0.55}{v1,v3,v2,v1}
\fmf{dbl_wiggly}{v1,v4,v3}
\fmf{dbl_wiggly}{v2,v4,v3}
\fmf{dbl_wiggly}{v1,v4,v2}
\end{fmfgraph}} ~+ \frac{1}{8}~
\parbox{20mm}{\begin{fmfgraph}(20,20)
\fmfforce{1mm,10mm}{v1}\fmfforce{19mm,10mm}{v2}
\fmfforce{10mm,19mm}{v3}\fmfforce{10mm,1mm}{v4}
\fmfcon{v1,v2}
\fmf{dbl_wiggly,right=0.5,tension=1}{v4,v2,v3,v1,v4}
\fmf{dbl_wiggly,right=0.5,tension=0.4}{v1,v3}
\fmf{dbl_wiggly,left=0.5,tension=0.4}{v2,v3}
\end{fmfgraph}} \\
&&-~ \frac{1}{3}~
\parbox{20mm}{\begin{fmfgraph}(20,20)
\fmfforce{10mm,10mm}{v4}\fmfforce{2.206mm,5.5mm}{v1}
\fmfforce{17.794mm,5.5mm}{v2}
\fmfforce{10mm,19mm}{v3}
\fmfcon{v1,v2,v3,v4}
\fmf{dbl_dots_arrow,left=0.6,tension=0.4}{v1,v3,v2,v1}
\fmf{dbl_wiggly}{v1,v4,v3}
\fmf{dbl_wiggly}{v2,v4,v3}
\fmf{dbl_wiggly}{v1,v4,v2}
\end{fmfgraph}} ~- \frac{1}{4}~
\parbox{20mm}{\begin{fmfgraph}(20,20)
\fmfforce{10mm,10mm}{v4}\fmfforce{2.206mm,5.5mm}{v1}
\fmfforce{17.794mm,5.5mm}{v2}
\fmfforce{10mm,19mm}{v3}
\fmfcon{v1,v2,v3,v4}
\fmf{dbl_dots_arrow,left=0.575,tension=0.4}{v1,v3,v2}
\fmf{dbl_wiggly,left=0.575,tension=0.4}{v2,v1}
\fmf{dbl_dots}{v4,v1}
\fmf{dbl_dots}{v2,v4}
\fmf{dbl_wiggly}{v3,v4}
\end{fmfgraph}}~ .
\end{eqnarray}
\end{fmffile}%
For an $n$-loop pure Yang-Mills planar diagram, the tracing over
internal color indices generically results in an overall color factor
of $(\Nc^2 - 1) \Nc^{n-1}$. Furthermore, since an $n$-loop vacuum
bubble is also proportional to $g^{2(n-1)}$, one finds as earlier that
factors of $g^2$ always appear in the form of the 't Hooft coupling
$g^2\Nc$. Hence, for the pure gauge problem, the natural scale is not
$g^2$, but rather $g^2 \Nc$. The Higgs contribution is
\begin{fmffile}{fmf_gammaH}
\begin{eqnarray}
\label{eq:GammaH}
\nonumber \Gamma_{\HIG} &=& \Tr \log D -
\Tr [D^{(0)}]^{-1} D \\
\nonumber &&+\frac{1}{2}~
\parbox{20mm}{\begin{fmfgraph}(20,20)
\fmfforce{1mm,0.5h}{v1}\fmfforce{19mm,0.5h}{v2}
\fmf{dbl_plain_arrow,left,tension=0.4}{v1,v2,v1}
\fmf{dbl_wiggly}{v1,v2}
\fmfcon{v1}
\end{fmfgraph}}
~+ \frac{1}{2}~
\parbox{20mm}{\begin{fmfgraph}(20,20)
\fmfforce{1mm,0.5h}{v1}\fmfforce{19mm,0.5h}{v3}
\fmfforce{0.5w,0.5h}{v2}
\fmf{dbl_plain,right=1,tension=0.4}{v1,v2}
\fmf{dbl_plain_arrow,right=1,tension=0.4}{v2,v1}
\fmf{dbl_wiggly,left,tension=0.4}{v3,v2,v3}
\end{fmfgraph}}
~+  ~
\parbox{20mm}{\begin{fmfgraph}(20,20)
\fmfforce{1mm,0.5h}{v1}\fmfforce{19mm,0.5h}{v3}
\fmfforce{0.5w,0.5h}{v2}
\fmf{dbl_plain,right=1,tension=0.4}{v1,v2}
\fmf{dbl_plain_arrow,right=1,tension=0.4}{v2,v1}
\fmf{dbl_plain,left=1,tension=0.4}{v3,v2}
\fmf{dbl_plain_arrow,left=1,tension=0.4}{v2,v3}
\end{fmfgraph}}\\
\nonumber &&+ \frac{1}{2}~
\parbox{20mm}{\begin{fmfgraph}(20,20)
\fmfforce{1mm,0.5h}{v1}\fmfforce{19mm,0.5h}{v2}
\fmf{dbl_plain_arrow,left=1,tension=0.4}{v1,v2}
\fmf{dbl_plain,left=1,tension=0.4}{v2,v1}
\fmf{dbl_plain_arrow,left=0.5}{v1,v2}
\fmf{dbl_plain,left=0.5}{v2,v1}
\end{fmfgraph}} ~+ \frac{1}{4}~
\parbox{20mm}{\begin{fmfgraph}(20,20)
\fmfforce{1mm,0.5h}{v1}\fmfforce{19mm,0.5h}{v2}
\fmf{dbl_plain_arrow,left,tension=0.4}{v1,v2,v1}
\fmf{dbl_wiggly,left=0.5}{v1,v2}
\fmf{dbl_wiggly,right=0.5}{v1,v2}
\end{fmfgraph}} ~+~
\parbox{20mm}{\begin{fmfgraph}(20,20)
\fmfforce{1mm,10mm}{v1}\fmfforce{19mm,10mm}{v2}
\fmfforce{10mm,19mm}{v3}\fmfforce{10mm,1mm}{v4}
\fmfcon{v1,v2}
\fmf{dbl_plain_arrow,right=1}{v1,v2}
\fmf{dbl_plain_arrow,right=0.5}{v2,v3}
\fmf{dbl_plain_arrow,right=0.5}{v3,v1}
\fmf{dbl_wiggly,right=0.5,tension=0.4}{v1,v3}
\fmf{dbl_wiggly,left=0.5,tension=0.4}{v2,v3}
\end{fmfgraph}} \\
&&+~ \frac{1}{3}~
\parbox{20mm}{\begin{fmfgraph}(20,20)
\fmfforce{10mm,10mm}{v4}\fmfforce{2.206mm,5.5mm}{v1}
\fmfforce{17.794mm,5.5mm}{v2}
\fmfforce{10mm,19mm}{v3}
\fmfcon{v1,v2,v3,v4}
\fmf{dbl_plain_arrow,left=0.6,tension=0.4}{v1,v3,v2,v1}
\fmf{dbl_wiggly}{v1,v4,v3}
\fmf{dbl_wiggly}{v2,v4,v3}
\fmf{dbl_wiggly}{v1,v4,v2}
\end{fmfgraph}} ~+ \frac{1}{4}~
\parbox{20mm}{\begin{fmfgraph}(20,20)
\fmfforce{10mm,10mm}{v4}\fmfforce{2.206mm,5.5mm}{v1}
\fmfforce{17.794mm,5.5mm}{v2}
\fmfforce{10mm,19mm}{v3}
\fmfcon{v1,v2,v3,v4}
\fmf{dbl_plain_arrow,left=0.575,tension=0.4}{v1,v3,v2}
\fmf{dbl_wiggly,left=0.575,tension=0.4}{v2,v1}
\fmf{dbl_plain}{v4,v1}
\fmf{dbl_plain}{v2,v4}
\fmf{dbl_wiggly}{v3,v4}
\end{fmfgraph}}
~ + ~~
\parbox{10mm}{\begin{fmfgraph}(10,20)
\fmfforce{0.5w,0.25h}{v1}\fmfforce{0.5w,0.75h}{v2}
\fmfv{decoration.shape=hexacross}{v2}
\fmf{dbl_plain_arrow,left=1}{v1,v2}
\fmf{dbl_plain,left=1}{v2,v1}
\end{fmfgraph}}
~~.
\end{eqnarray}
\end{fmffile}%
These diagrams have a somewhat more complicated dependence on $\Nc$
(the associated color factors are stated in Table
\ref{Higgstopologies}). We will nevertheless continue to use $g^2 \Nc$
as the standard mass scale, but for clarity, units of $g^2\Nc$ will be
explicitly stated throughout.
\begin{fmffile}{fmf_2looptopologies}
\begin{table}
\centering
\begin{tabular}{c@{\hspace{5mm}}c@{\hspace{5mm}}c@{\hspace{5mm}}c@{\hspace{5mm}}c}
\parbox{20mm}{\begin{fmfgraph}(20,20)
\fmfforce{1mm,0.5h}{v1}\fmfforce{19mm,0.5h}{v2}
\fmfcon{v1,v2}
\fmf{dbl_plain_arrow,left,tension=0.4}{v1,v2,v1}
\fmf{dbl_wiggly}{v1,v2}
\end{fmfgraph}}
&
\parbox{20mm}{\begin{fmfgraph}(20,20)
\fmfforce{1mm,0.5h}{v1}\fmfforce{19mm,0.5h}{v3}
\fmfforce{0.5w,0.5h}{v2}
\fmf{dbl_plain,right=1,tension=0.4}{v1,v2}
\fmf{dbl_plain_arrow,right=1,tension=0.4}{v2,v1}
\fmf{dbl_wiggly,left,tension=0.4}{v3,v2,v3}
\end{fmfgraph}}
&
\parbox{20mm}{\begin{fmfgraph}(20,20)
\fmfforce{1mm,0.5h}{v1}\fmfforce{19mm,0.5h}{v3}
\fmfforce{0.5w,0.5h}{v2}
\fmf{dbl_plain,right=1,tension=0.4}{v1,v2}
\fmf{dbl_plain_arrow,right=1,tension=0.4}{v2,v1}
\fmf{dbl_plain,left=1,tension=0.4}{v3,v2}
\fmf{dbl_plain_arrow,left=1,tension=0.4}{v2,v3}
\end{fmfgraph}}
&& \vspace{3mm}\\ \vspace{3mm}
$a$ & $b$ & $c$  & &\\
\parbox{20mm}{\begin{fmfgraph}(20,20)
\fmfforce{10mm,10mm}{v4}\fmfforce{2.206mm,5.5mm}{v1}
\fmfforce{17.794mm,5.5mm}{v2}
\fmfforce{10mm,19mm}{v3}
\fmfcon{v1,v2,v3,v4}
\fmf{dbl_plain_arrow,left=0.575,tension=0.4}{v1,v3,v2}
\fmf{dbl_wiggly,left=0.575,tension=0.4}{v2,v1}
\fmf{dbl_plain}{v4,v1}
\fmf{dbl_plain}{v2,v4}
\fmf{dbl_wiggly}{v3,v4}
\end{fmfgraph}}
&
\parbox{20mm}{\begin{fmfgraph}(20,20)
\fmfforce{10mm,10mm}{v4}\fmfforce{2.206mm,5.5mm}{v1}
\fmfforce{17.794mm,5.5mm}{v2}
\fmfforce{10mm,19mm}{v3}
\fmfcon{v1,v2,v3,v4}
\fmf{dbl_plain_arrow,left=0.575,tension=0.4}{v1,v3,v2}
\fmf{dbl_plain_arrow,left=0.575,tension=0.4}{v2,v1}
\fmf{dbl_wiggly}{v4,v1}
\fmf{dbl_wiggly}{v2,v4}
\fmf{dbl_wiggly}{v3,v4}
\end{fmfgraph}}
&
\parbox{20mm}{\begin{fmfgraph}(20,20)
\fmfforce{1mm,10mm}{v1}\fmfforce{19mm,10mm}{v2}
\fmfforce{10mm,19mm}{v3}\fmfforce{10mm,1mm}{v4}
\fmfcon{v1,v2}
\fmf{dbl_plain_arrow,right=1}{v1,v2}
\fmf{dbl_plain_arrow,right=0.5}{v2,v3}
\fmf{dbl_plain_arrow,right=0.5}{v3,v1}
\fmf{dbl_wiggly,right=0.5,tension=0.4}{v1,v3}
\fmf{dbl_wiggly,left=0.5,tension=0.4}{v2,v3}
\end{fmfgraph}}
&\parbox{20mm}{\begin{fmfgraph}(20,20)
\fmfforce{1mm,0.5h}{v1}\fmfforce{19mm,0.5h}{v2}
\fmf{dbl_plain_arrow,left=1,tension=0.4}{v1,v2}
\fmf{dbl_plain,left=1,tension=0.4}{v2,v1}
\fmf{dbl_plain_arrow,left=0.5}{v1,v2}
\fmf{dbl_plain,left=0.5}{v2,v1}
\end{fmfgraph}}
&
\parbox{20mm}{\begin{fmfgraph}(20,20)
\fmfforce{1mm,0.5h}{v1}\fmfforce{19mm,0.5h}{v2}
\fmf{dbl_plain_arrow,left,tension=0.4}{v1,v2,v1}
\fmf{dbl_wiggly,left=0.5}{v1,v2}
\fmf{dbl_wiggly,right=0.5}{v1,v2}
\end{fmfgraph}}
\vspace{3mm}\\
$A$ & $B$ & $C$ & $D$ & $E$
\vspace{5mm}
\end{tabular}
\begin{tabular}{l@{\hspace{5mm}}l}
$(a)$ & $d_AT_R$ \\
$(b)$ & $2d_AT_R$ \\
$(c)$ & $d_R (1+d_R)$ \\
$(A)$ & $d_A T_R ( C_R - \frac{1}{2}C_A)$ \\
$(B)$ & $\frac{1}{2} d_A T_R C_A$ \\
$(C)$ & $d_A T_R ( 2C_R - \frac{1}{2}C_A)$ \\
$(D)$ & $2d_R (1+d_R)$ \\
$(E)$ & $d_A T_R ( 4C_R - C_A)$
\end{tabular}
\caption{\label{Higgstopologies}Color factors for the two and three-loop Higgs topologies.}
\end{table}\end{fmffile}

The power of the 2PI formalism becomes apparent when we perform the variation of $\Gamma$ with respect to $\GT$, $\GL$, $\Delta$ and $D$. For instance, from $\delta \Gamma/ \delta D = 0$, we have
\begin{fmffile}{fmf_SDequation}
\be \label{eq:GSD}
-D^{-1}(p) + D^{(0) -1}(p) = \Pi_\HIG(p)
\ee
with (omitting charge arrows)
\bea
\nonumber \Pi_\HIG(p) &=& ~
\parbox{20mm}{\begin{fmfgraph}(20,20)
\fmfleft{l1}\fmfright{r1}
\fmfforce{0.25w,0.5h}{v1}\fmfforce{0.75w,0.5h}{v3}
\fmf{dbl_plain}{l1,v1}\fmf{dbl_plain,right=1}{v1,v3}\fmf{dbl_wiggly,right=1}{v3,v1}\fmf{dbl_plain}{v3,r1}
\end{fmfgraph}}
~ + 2 ~
\parbox{20mm}{\begin{fmfgraph}(20,20)
\fmfleft{l1}\fmfright{r1}\fmftop{t1}\fmfforce{0.5w,0.5h}{v1}
\fmf{dbl_plain}{l1,v1}\fmf{phantom,tension=5.0}{t1,v2}
\fmf{dbl_plain,left=1,tension=0.4}{v1,v2,v1}
\fmf{dbl_plain}{v1,r1}
\end{fmfgraph}}
~ + \frac{1}{2} ~
\parbox{20mm}{\begin{fmfgraph}(20,20)
\fmfleft{l1}\fmfright{r1}\fmftop{t1}\fmfforce{0.5w,0.5h}{v1}
\fmf{dbl_plain}{l1,v1}\fmf{phantom,tension=5.0}{t1,v2}
\fmf{dbl_wiggly,left=1,tension=0.4}{v1,v2,v1}
\fmf{dbl_plain}{v1,r1}
\end{fmfgraph}}\\
\nonumber && + ~
\parbox{20mm}{\begin{fmfgraph}(20,20)
\fmfleft{l1}\fmfright{r1}\fmfforce{0.25w,0.5h}{v1}\fmfforce{0.75w,0.5h}{v2}
\fmfforce{0.5w,0.75h}{vt}\fmfforce{0.5w,0.25h}{vb}
\fmf{dbl_plain}{l1,v1}
\fmf{dbl_wiggly,right=0.5}{v1,vb}
\fmf{dbl_wiggly,left=0.5}{vt,v2}
\fmf{dbl_plain,left=0.5}{v1,vt}
\fmf{dbl_plain,right=0.5}{vb,v2}
\fmf{dbl_plain}{v2,r1}\fmf{dbl_plain}{vt,vb}
\end{fmfgraph}}
~ + ~
\parbox{20mm}{\begin{fmfgraph}(20,20)
\fmfleft{l1}\fmfright{r1}\fmfforce{0.25w,0.5h}{v1}\fmfforce{0.75w,0.5h}{v2}
\fmfforce{0.5w,0.75h}{vt}\fmfforce{0.5w,0.25h}{vb}
\fmf{dbl_plain}{l1,v1}
\fmf{dbl_wiggly,left=0.5}{v1,vt,v2}\fmf{dbl_wiggly}{vt,vb}
\fmf{dbl_plain,right=0.5}{v1,vb}
\fmf{dbl_plain,right=0.5}{vb,v2}
\fmf{dbl_plain}{v2,r1}
\end{fmfgraph}}
~ + 2 ~
\parbox{20mm}{\begin{fmfgraph}(20,20)
\fmfleft{l1}\fmfright{r1}\fmfforce{0.25w,0.5h}{v1}\fmfforce{0.75w,0.5h}{v2}
\fmfforce{0.5w,0.75h}{vt}\fmfforce{0.5w,0.25h}{vb}
\fmf{dbl_plain}{l1,v1}
\fmf{dbl_wiggly,right=0.5}{vt,v2}
\fmf{dbl_plain,left=0.5}{v1,vt}
\fmf{dbl_plain,left=0.5}{vt,v2}
\fmf{dbl_wiggly,left=0.5}{v2,vb,v1}
\fmf{dbl_plain}{v2,r1}
\end{fmfgraph}}\\
&&+ ~ 2 ~
\parbox{20mm}{\begin{fmfgraph}(20,20)
\fmfleft{l1}\fmfright{r1}\fmfforce{0.25w,0.5h}{v1}\fmfforce{0.75w,0.5h}{v2}
\fmf{dbl_plain}{l1,v1}\fmf{dbl_plain,left=1}{v1,v2,v1}
\fmf{dbl_plain}{v1,v2}
\fmf{dbl_plain}{v2,r1}
\end{fmfgraph}}
~ + \frac{1}{2} ~
\parbox{20mm}{\begin{fmfgraph}(20,20)
\fmfleft{l1}\fmfright{r1}\fmfforce{0.25w,0.5h}{v1}\fmfforce{0.75w,0.5h}{v2}
\fmf{dbl_plain}{l1,v1}
\fmf{dbl_plain}{v1,v2}
\fmf{dbl_plain}{v2,r1}\fmf{dbl_wiggly,left=1}{v1,v2,v1}
\end{fmfgraph}}
~ + ~
\parbox{20mm}{\begin{fmfgraph}(20,20)
\fmfleft{l1}\fmfright{r1}\fmfforce{0.25w,0.5h}{v1}\fmfforce{0.75w,0.5h}{v3}
\fmfforce{0.5w,0.5h}{v2}
\fmf{dbl_plain}{l1,v1}
\fmf{dbl_plain,left=1}{v1,v2}
\fmf{dbl_wiggly,left=1}{v2,v1}
\fmf{dbl_plain}{v3,r1}
\fmf{dbl_plain,left=1}{v2,v3}
\fmf{dbl_wiggly,left=1}{v3,v2}
\end{fmfgraph}} ~+~
\parbox{20mm}{\begin{fmfgraph}(20,20)
\fmfleft{l1}\fmfright{r1}
\fmfforce{0.5w,0.5h}{v1}
\fmfv{decoration.shape=hexacross}{v1}
\fmf{dbl_plain}{l1,v1,r1}
\end{fmfgraph}} ~. \qquad \label{eq:PiphiSD}
\eea
\end{fmffile}
Equations of the type \Eq{eq:GSD} / \Eq{eq:PiphiSD} are
generically referred to in this work as SD equations, and the
topologies which appear in \Eq{eq:PiphiSD} correspond to the loop
order of the truncation of the effective action. By solving this
equation self-consistently in a three-loop truncation, we fully resum
one- and two-loop self-energy topologies to all orders.

\Eq{eq:PiphiSD} contains terms that are linearly and logarithmically
divergent; in dimensional regularization, only the logarithmic
divergences appear explicitly as $1/\epsilon$'s, and these are
subtracted by the counterterm. This implies that the entire
computation must be performed in $\MSbar$, which requires the analytic
continuation of these integrals to $\Dim$ dimensions. The
regularization procedure which we adopt is described at length in
\cite{3PI}; to quickly recap the key points, consider the tadpole
graph
\begin{fmffile}{fmf_regularization}
\be
\mathcal{A} = -\frac{1}{2 \lambda(d_R + 1)}~\parbox{20mm}{\begin{fmfgraph}(20,20)
\fmfleft{l1}\fmfright{r1}\fmftop{t1}\fmfforce{0.5w,0.5h}{v1}
\fmf{dbl_plain}{l1,v1}\fmf{phantom,tension=5.0}{t1,v2}
\fmf{dbl_plain,left=1,tension=0.4}{v1,v2,v1}
\fmf{dbl_plain}{v1,r1}
\end{fmfgraph}} ~= \frac{1}{{\bar \mu}^{2\epsilon}}\int \frac{d^{\Dim} q}{(2\pi)^{\Dim}} D(q)
\ee
\end{fmffile}
with $\bar\mu^2 = \mu^2 e^\gamma / 4\pi$, and $\Dim =
3+2\epsilon$. Since $D(q)$ is an arbitrary function of $q$, this
integral would need to be performed numerically; in doing so we must
set $\Dim \rightarrow 3$. To implement dimensional regularization, we
adopt a procedure of ``addition and subtraction,'' as follows:
\be
\mathcal{A} = \frac{1}{{\bar \mu}^{2\epsilon}}\int \frac{d^{\Dim}
  q}{(2\pi)^{\Dim}} \Big ( D(q) - \frac{1}{q^2} \Big )  +
\frac{1}{{\bar \mu}^{2\epsilon}}\int \frac{d^{\Dim} q}{(2\pi)^{\Dim}}
\frac{1}{q^2}.
\ee
The rightmost term is simple enough that it can be computed
analytically (in $\MSbar$ its value is zero), and the leftmost term is
now only logarithmically divergent. Thus, we have removed the linear
divergence by subtracting $1/q^2$, and now the next step is to remove
the logarithmic one. At large momenta, and near 3 dimensions, $D(q)$
can be expanded as
\be
D(q) \sim \frac{1}{q^2} + \frac{g^2 C_R \big(1 + \epsilon (1 - \xi -
  \log 4)\big)}{4 \mu^{2\epsilon} q^{3-2\epsilon}} + \mathcal{O}\Big (
\frac{1}{q^4} \Big )
\ee
where we have been careful to keep $\mathcal{O}(\epsilon)$ corrections
in the $1/q^3$ term. Now, we can add and subtract the subleading term,
\bea
&&\mathcal{A} =\frac{1}{\bar \mu^{2\epsilon}}\int \frac{d^{\Dim}
  q}{(2\pi)^{\Dim}} \Bigg [ D(q) - \frac{1}{q^2} - \frac{g^2 C_R
    \big(1 + \epsilon (1 - \xi + \log 4)\big)}{4\mu^{2\epsilon}(q^2 +
    \omega^2)^{3/2-\epsilon}}\Bigg ] \nonumber\\
&&\quad + \frac{1}{\bar \mu^{2\epsilon}}\int \frac{d^{\Dim}
  q}{(2\pi)^{\Dim}} \frac{1}{q^2} + \frac{g^2 C_R \big(1 + \epsilon (1
  - \xi - \log 4)\big)}{4 \bar \mu^{2\epsilon} \mu^{2\epsilon}}\int
\frac{d^{\Dim}
  q}{(2\pi)^{\Dim}}\frac{1}{(q^2+\omega^2)^{3/2-\epsilon}}. \qquad
~\label{eq:Areg}
\eea
The first line of \Eq{eq:Areg} is finite, so we can set $\Dim=3$ and
perform the integral numerically. What we have effectively done is
shuffled all of the $\epsilon$ dependence into terms which can be
integrated analytically. Thus the regularized expression for
$\mathcal{A}$ has the form
\bea
\mathcal{A} &=& \int \frac{d^{3} q}{(2\pi)^{3}} \Bigg [ D(q) -
  \frac{1}{q^2} - \frac{g^2 C_R}{4(q^2 + \omega^2)^{3/2}}\Bigg ]
\nonumber\\
&&+~ \frac{g^2 C_R \big(1 + \epsilon (1 - \xi - \log 4)\big)}{4
  (4\pi)^{3/2} e^{\epsilon \gamma}
  \Gamma(3/2-\epsilon)}\Gamma(-2\epsilon) \Big (
\frac{\omega}{\mu}\Big )^{4\epsilon} .
\eea
We can then subtract the $1/\epsilon$ divergence with the
counterterm, and take the limit $\epsilon \rightarrow 0$. Note here
that the subleading term appears with a mass $\omega$. Its value is
arbitrary, but it must be included, otherwise one would introduce an
IR divergence where originally there was none. For simplicity, we can
set $\omega = g^2\Nc$ noting that the final results of the calculation
are $\omega$ independent. Though it is certainly permitted, it is not
a requirement that $\omega$ be set to the scalar mass $m$ (and our
reasoning for not doing so is explained in Appendix
\ref{SelfEnergy}).

Other diagrams which appear in $\Gamma$ are regularized in much the
same fashion. In the end we need to compute all of the one- and
two-loop gluon and Higgs self-energy diagrams which appear in
perturbation theory (ensuring that IR divergent diagrams are not
introduced  inadvertently); the results of this computation are
contained in Appendix \ref{SelfEnergy}.

\subsection{Remarks on gauge fixing}

\label{gaugefix}

Since our approach is diagrammatic and is founded on trying to determine
correlation functions of gauge dependent fields, we are obliged to
perform some sort of gauge fixing.  We have chosen covariant gauge with
gauge-fixing functional $\partial_\mu A^\mu$, that is, a gauge choice
which does not make reference to the scalar field one-point function or
vacuum value.  This choice differs from what is usually done in
perturbative treatments of the Higgs phase, and requires some
explanation.  First we will argue that one \textsl{can} gauge fix as we
do here; then we will explain why we believe it is preferable.

That the gauge-fixing approach we have adopted is possible, has already
been explained clearly by Buchm\"uller, Fodor, and Hebecker
\cite{Buchmuller1} in the context of the electroweak phase transition.
Perturbatively, in the broken symmetry phase we
expect the typical contribution to the path integral to have a
nonvanishing scalar VEV $\Phi_i$; but since the gauge fixing does not
eliminate the integration over the global gauge rotations, there are
equal contributions from every field direction choice, and the ensemble
average of $\Phi_i$ is zero.  However the existence of a VEV will still
appear as a delta-function contribution to the scalar two-point
function, so the approach will still capture that physics.
Nonperturbatively, while infrared gauge fields are suppressed, we do not
expect them to vanish, and they can still destroy any infinite-range
order in the scalar field.  If this is the case then the scalar
two-point function will not in fact have a strict delta-function
contribution.  Instead it will have a very sharp structure near zero
momentum, corresponding to long (but not infinite) distance correlations
in the scalar field.  Indeed, we expect this must be the correct
behavior, since the symmetric and Higgs phases are analytically
connected and are not distinct in the sense of being distinguished by a
true order parameter.  But the existence of infinite-range Higgs-field
correlations in part but not all of the phase diagram would constitute
an order parameter and would forbid an analytic connection between the
symmetric and Higgs sides of the transition line.

Now let us consider the alternative approach.  It is to include
explicitly a one-point source for the scalar field,
\be
W[J_i,K_{ij}] =  - \log \int D[\phi,A] ~ e^{-S - J_i \phi_i -
  \frac{1}{2}\phi_i K_{ij} \phi_j} \,.
\ee
The value of $W[0,0]$ is gauge-invariant \cite{Fukuda1}, but the
inclusion of nonzero $J_i$ explicitly breaks gauge invariance.
Naturally we are then only interested in the $J_i \to 0$ limit.
Depending on our gauge-fixing procedure, this limit may or may not be
smooth.  That is, we can interpret the gauge-fixed one-point function as
a directional derivative
\be
\langle \Phi_i \rangle = \frac{\delta}{\delta J_i}
   W[J_i,0]\Big \vert_{J_i \rightarrow 0(\vartheta)}
\ee
where $0(\vartheta)$ means ``zero is approached along a direction
$\vartheta$ on the manifold of $\SUN$ rotations.''

Perturbatively, we expect the symmetric phase to display smooth
behavior, so an approach from any direction will yield the same result,
consistent with a zero VEV.  In the broken phase, $W$ is expected to
develop a conical singularity, so the direction has a significance.
Nonperturbatively, and still working in covariant gauge, we have just
argued that we do not in fact expect infinite-range correlations in the
scalar field when $J_i$ is taken to zero; so the behavior of $W$ near
$J_i=0$ should always be differentiable, albeit the derivatives can
become very large.  Therefore, in covariant gauge we expect that there
should be no VEV, even if $J_i$ is taken to zero along a particular
direction.  Therefore the introduction of $J_i$ changes nothing; the
$J_i \to 0$ limit is nonsingular and the VEV of the field vanishes.

Alternatively, we can change our gauge-fixing procedure so that it makes
explicit reference to the VEV $\Phi_i$ -- that is, we can use $R_\xi$
gauge.  The gauge-fixing choice introduces into the action the
gauge-fixing term
\be
\mathcal{L}_{gf} = \frac{1}{\xi}
 \left( \partial_\mu A^\mu_a - \xi \Phi_i g T^a_{ij} \phi_j \right)^2
 \,,
\ee
where $\Phi_i$ is the VEV and $\phi_j$ is the field.  This is balanced
as usual by the appropriate Fadeev-Popov determinant.  Physically, the
role of covariant gauge fixing can be understood as using up the gauge
freedom to force $\partial_\mu A^\mu$ to be as small as possible, which
minimizes the total size of fluctuations in the gauge fields.
$R_\xi$ gauge is instead a compromise, in which the gauge fixing is used
to try to minimize $\partial_\mu A^\mu$ (gauge field fluctuations), but
also to minimize fluctuations in the components of the scalar field not
in the direction of the VEV $\Phi_i$ (pseudo-Goldstone fluctuations).
The limit $\xi\to 0$, Landau gauge, is when all gauge freedom is used to
control gauge field fluctuations.  The opposite limit, $\xi\to \infty$,
Unitary gauge, is when all freedom is used to align the scalar in the
direction of its VEV.

One challenge with this approach is that in the current context the
value of the VEV $\Phi_i$ must be determined self-consistently as part
of the procedure.  In general the VEV will depend on $\xi$
\cite{Nielsen1}, growing
larger at large $\xi$ as more fluctuations are forced into the VEV.
Perturbatively this effect is suppressed by $g^2$ in 4 dimensions.  Here
it will be suppressed by $x$.  Since we are interested in a regime where
perturbation theory requires resummation, the $\xi$ dependence can be
large.

The first problem with this approach is that whether $\Phi_i$ vanishes
or not would constitute an order parameter, but we know that there
should not be an order parameter for this system.  Second, it is
possible that there are multiple self-consistent solutions for $\Phi_i$,
in which case it is not clear which to use. Finally, the existence of a
VEV $\Phi_i$ for a given $x,y$ value can and almost surely will depend
on $\xi$, as large $\xi$ biases the gauge fixing towards the development
of a VEV.  Therefore we anticipate that the details of the transition
will have no stability as a function of $\xi$, in other words, the
methodology would not be reliable.  (Similar issues are discussed in
Appendix A of Ref.~\cite{Arnold1}.)

On the other hand, we emphasize that using covariant gauge will present
considerable challenges when the transition is strong or when the value
of $y$ places us deep in the Higgs phase.  In this case we will be
keeping track of very long-distance correlations in the $\phi$ field via
the corresponding very low-momentum structure in the two-point
function.  As we will explain in the following, this proves numerically
challenging, but it can also be a problem from the point of view of the
convergence of the loop expansion and the stability of the solutions we
find within the space of possible $G,D,\Delta$ choices.  It is also not
clear what $\xi$ dependence the phase diagram will display in covariant
gauge; if the dependence is strong, it indicates a problem with the
method's reliability.

\section{\label{Extremization}Extremization of the effective action}

\subsection{Variational \ansatze}

In \cite{3PI}, we extensively described an algorithm which can be used
to extremize the effective action when only gauge fields are
present. Now we have to address the additional complications which arise
due to the presence of a Higgs field. In the symmetric phase, the
presence of the Higgs does not really change much at the technical
level, and obtaining self-consistent solutions for the gauge field and
Higgs propagators proceeds much as earlier.

To begin, we will review the details of the functions which enter into
the problem. Since we have assumed a general covariant gauge, we are
attempting to solve self-consistently for the following 4 functions:
$\GT(p)$, $\GL(p)$, $\Delta(p)$ and $D(p)$, which are respectively the
transverse and longitudinal gauge field propagators, the ghost
propagator and the Higgs propagator. We can opt for the most part to
simplify the problem further by working in Landau gauge, where $\GL$
falls out of the picture; however, computations in Feynman gauge do
require a treatment of $\GL$.

To realize the extremization, we will specify \textit{Ans\"atze} for
these functions in terms of a finite set $\mathbb{C} = \{c_i\}$ of
variational coefficients, such that the variational equations
become
\be
\frac{\delta\Gamma}{\delta \{\GT / \GL / \Delta / D\}} = 0
\qquad \longrightarrow \qquad
\frac{\delta \Gamma}{ \delta c_i} = 0 \,.
\ee
Writing the functions in terms of a finite number of parameters in this
way replaces the infinite-dimensional functional space with a
finite-dimensional subspace; and the problem becomes finding the
extremum in this subspace.
By increasing the size of $\mathbb{C}$ we enlarge the space of
allowed functions, and the true extremum should be more closely
approached.  Our choice is to fit the self-energies as rational
functions (Pad\'e approximants), since this gives a very flexible class
of smooth functions.  Specifically, for some $\{a_i\}\cup
\{b_j\} \subset \mathbb{C}$, we will define
\be
\mathcal{R}_{i_{\text{max}} - j_{\text{max}}}(p,\{a_i\}\cup \{b_j\}) =
\frac{a_{i_{\text{max}}}p^{i_{\text{max}}} + ... +
  a_0}{b_{j_{\text{max}}}p^{j_{\text{max}}} + g^2\Nc}.
\ee
Then the two-point
functions \Eq{eq:Dfull}, \Eq{eq:Deltafull} and \Eq{eq:Gfull} are
parametrized as follows:
\bea
 \label{eq:DT} \Pi_T(p) &=& g^2 \Big (\frac{\Nc(\xi^2 + 2\xi + 11)}{64}
 -\frac{T_R}{ 16}\Big ) p + \mathcal{R}_0(p,\{c^{\{\Pi_T\}}_i\})  \\
\label{eq:DL} \Pi_L(p) &=& \mathcal{R}_0(p,\{c^{\{\Pi_L\}}_i\})\\
\label{eq:Delta} \Sigma(p) &=& \frac{g^2\Nc}{16}
\frac{p^2}{p+ g^2 \Nc } + \mathcal{R}_0(p,\{c^{\{\Sigma\}}_i\})\\
\label{eq:G} \Pi_{\HIG}(p) &=&  \frac{g^2 C_R}{ 4}
\frac{p^2}{p+ g^2 \Nc} +
\mathcal{R}_0(p,\{c^{\{\Pi_{\HIG}\}}_i\}) .
\eea
Here we have incorporated the one-loop large-$p$ behavior exactly and
allowed the rest of the self-energy to be determined by extremization.
Previous work \cite{3PI} shows that third order Pad\'e
approximants are sufficient and we will use them here.
The resulting SD equations have the simple form
$\Pi^{\text{\textit{Ansatz}}}_T (p) =
\Pi^{\text{2PI}}_T(p)$%
\footnote{
   Using the labels ``\ansatz '' and
   ``2PI'' to distinguish between the value of the Pad\'e approximant and
   self-energy functional constructed out of 2PI diagrams.}
(where
$\Pi^{\text{2PI}}_T(p)$ is the gluonic analogue of \Eq{eq:PiphiSD}) and
similarly for $\Pi_L$, $\Sigma$ and $\Pi_{\HIG}$.

However, we anticipate that the scalar propagator may display a very
narrow structure near zero momentum.  Therefore we will add to the
scalar propagator \ansatz\ an additional term:
\be
\label{eq:Gfullansatz}
D(p) = \frac{\mathcal{R}_0(p,\{c_i^{\{G\}}\})}{p^{2}(p^{2}+
  g^4 N^2)} + \frac{1}{p^2 + m^2 - \Pi_{\HIG}(p)} .
\ee
The added term is designed to allow for a sharp structure at small $p$;
its form has been chosen phenomenologically.  Technically the functional
form allows for $D(p) \propto p^{-2}$ small-$p$ behavior, whereas we
expect that $\lim_{p\to 0}D(p)$ should be a constant.%
\footnote{
    Generally we expect self-energies to be nonzero at $p=0$ and so
    propagators should go to constant values.  The exception is the
    ghost propagator, where we showed in \cite{3PI} that the self-energy
    must scale as $\Sigma(p) \propto p^2$ at small $p$ because a vertex
    always differentiates the external ghost line.
}
However, the extremization procedure is not sensitive to a turnover at
very small $p$, so this functional form near $p=0$ is not very
important.  Generally, in the
symmetric phase extremization setting $\mathcal{R}_0$ to zero produces
essentially the same extremum as allowing the term to be nonzero,
whereas deep in the Higgs region the inclusion of
$\mathcal{R}_0$ is essential to getting a good solution to the SD
equation, which is modified to
$-D^{-1}(p) + D^{(0) -1}(p) = \Pi_{\HIG}^{\text{2PI}}(p)$.

We can also define the renormalized scalar ``condensate'' as
\bea
\label{eq:curlyG}
\Dcond &=& \int \frac{d^3 q}{(2\pi)^3} \Bigg [D(q) -
\frac{1}{q^2} - \frac{g^2C_R}{4(q^2 + g^4\Nc^2)^{3/2}} \Bigg ]
\\
& = & \frac{1}{\Nc}\langle \phi^\dagger \phi \rangle_{\bar\mu=g^2 N} \,.
\nonumber
\eea
The twice-subtracted integral is both UV and IR finite;
as indicated it equals the expectation value of the field
squared when renormalized using $\bar\mu = g^2 N$.  This condensate will
be useful in distinguishing between coexisting phases.

In \Eq{eq:DT}, \Eq{eq:Delta}, and \Eq{eq:G} we have fixed by hand the
$\mathcal{O}(p)$ large-momentum behavior of each propagator to match a
one-loop perturbative
calculation presented in Appendix \ref{SelfEnergy}, leaving only the
$\mathcal{O}(p^0)$ part to be determined variationally.
This is actually a requirement; to see why this is the case,
consider the UV expansion of $\GT$ resulting from \Eq{eq:DT},
\be
\label{eq:DTUV}\GT(p \gg g^2\Nc) = \frac{1}{p^2} + \frac{\frac{g^2\Nc
    (\xi^2 + 2\xi + 11)}{64} -\frac{g^2T_R}{16}}{p^3} + \mathcal{O}\Big
( \frac{1}{p^4}\Big ) ,
\ee
as well as the variation of $\Gamma$ with respect to $c_i^{\{\Pi_T\}}$
\be
 \label{eq:DTSDequation}
\frac{\delta \Gamma}{\delta c_i^{\{\Pi_T\}}} = d_A \int \frac{d^3
  p}{(2\pi)^3} \frac{\delta\GT(p)}{\delta c_i^{\{\Pi_T\}}}\Big (
-\Pi^{\text{\textit{Ansatz}}}_T(p) + \Pi^{\text{2PI}}_T(p)\Big) .
\ee
By fixing the tree-level $\mathcal{O}(1/p^2)$ and one-loop
$\mathcal{O}(1/p^3)$ behavior in \Eq{eq:DTUV}, the term in parentheses
in \Eq{eq:DTSDequation} is automatically $\mathcal{O}(1)$ at large
momentum, while the derivative of $\GT$ is $\mathcal{O}(1/p^4)$ (or
milder, depending on which coefficient we are differentiating with
respect to). Hence, \Eq{eq:DTSDequation} is finite. Finally, it is worth
noting that at one loop the inclusion of masses in bare diagrams is
subleading in $p$ relative to the massless diagrams; as we are not
required to impose any constraints on the two-point functions at
$\mathcal{O}(1/p^4)$, it suffices to compute the one-loop corrections in
the massless limit.

Since convergence to the perturbative limit is only necessary at large
$p$, looking back at  \Eq{eq:Delta} and \Eq{eq:G}, we opted to include
the one-loop contributions with an additional IR suppression factor of
$p/(p+\omega)$. At sufficiently small momenta, a
linear term in the denominator of a propagator can lead to the formation
of a pole. The gauge fields dynamically generate a mass that is
sufficiently large to prevent this sort of thing from happening so a
suppression factor of this sort is not required. For the Higgs and ghost
this is not the case. In solving the problem we
have set $\omega = g^2\Nc$.  This arbitrary
choice does not affect the final result, since a different
choice of $\omega$ together with an appropriate shift in the
\ansatz\ parameters leaves the self-energy unchanged.

\subsection{Initial conditions and root finding}

To extremize $\Gamma[\GT,\GL,\Delta,D]$, we employ and algorithm based
on conjugate gradient descent specialized to the problem at hand. We can
visualize the root-finding algorithm as a dynamical system where we
choose an initial value for the coefficients $c_i$ and subsequently
follow a flow through the gradient field $\partial \Gamma / \partial
c_i$ until we reach an attracting fixed point, which corresponds to a
solution.

For $x$ below the critical end point, there is a region of metastability
where two attractors coexist.  We will denote the solution with
larger condensate $\Dcond$ (and smaller small-$p$ behavior in
$G_T$) as the Higgs solution and the solution with smaller
condensate as the symmetric phase solution.  We write the propagator (condensate) in
the Higgs solution as $\Dhig$ ($\Dcond_-$) and that in the symmetric phase as $\Dsym$ ($\Dcond_+$).
In the space of initial guesses for the \ansatz\ parameters, each
solution has a basin of attraction, which we write as $\gamma_{+}$ and
$\gamma_{-}$.  These basins of attraction do not cover the full space of
initial guesses; because of the saddle-like nature of $\Gamma$, there
is also a set of initial conditions $c_i\in\gamma_0$
which evolve towards divergent values of $D$. This is to be discussed in
greater detail in Section \ref{Results}.

\begin{figure}
\centering
\includegraphics[scale=0.75]{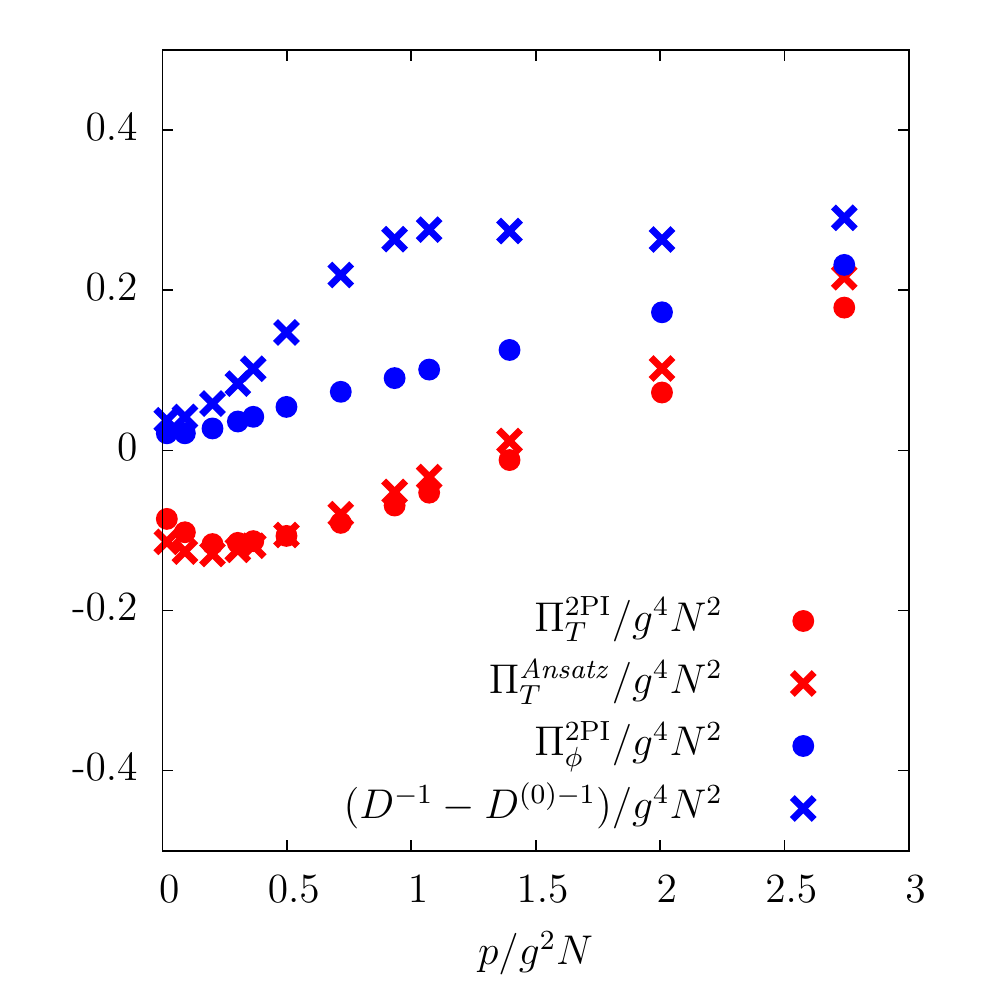}
\includegraphics[scale=0.75]{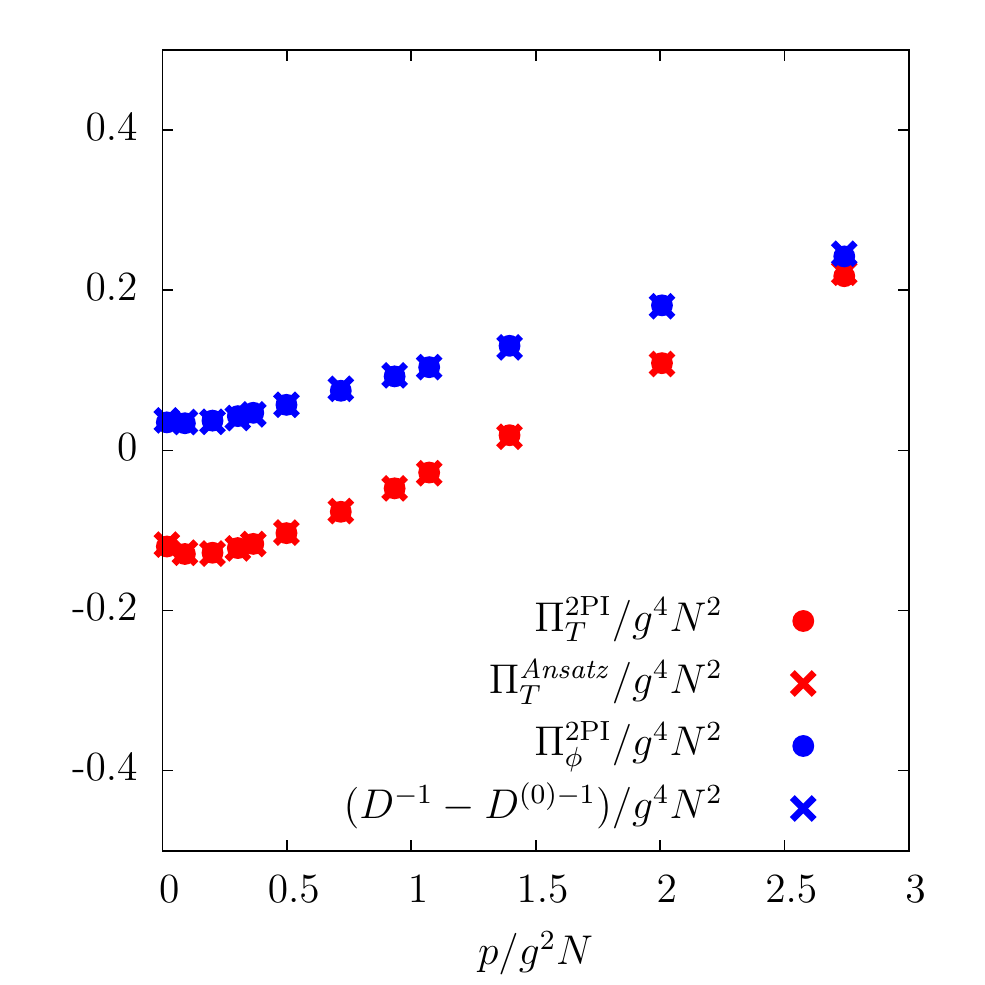}
\caption{\label{fig:Higgsconvergence}Evolution of the SD equations under
  gradient descent, which are solved when the points overlap in the
  above figures (the ghost equation is not depicted, but it is
  qualitatively similar). The left panel corresponds to some initial
  choice of variational coefficients, and at the right we see
  convergence at late times.}
\end{figure}

The number of iterations of gradient descent in the extremization
procedure (denoted by $\mathcal{N}$) can be thought of as time
evolution, and we are interested in the results at late times. We can
observe convergence of the algorithm by plotting the evolution of the
LHS and RHS of the SD equations with $\mathcal{N}$; this is shown
generically in Fig.~\ref{fig:Higgsconvergence}. {}From this figure it is
apparent that convergence is attained, and that the variational \ansatz\
has captured a choice for the self-energy where the SD equations are
quite accurately solved.

\begin{figure}
\centering
\includegraphics[scale=0.75]{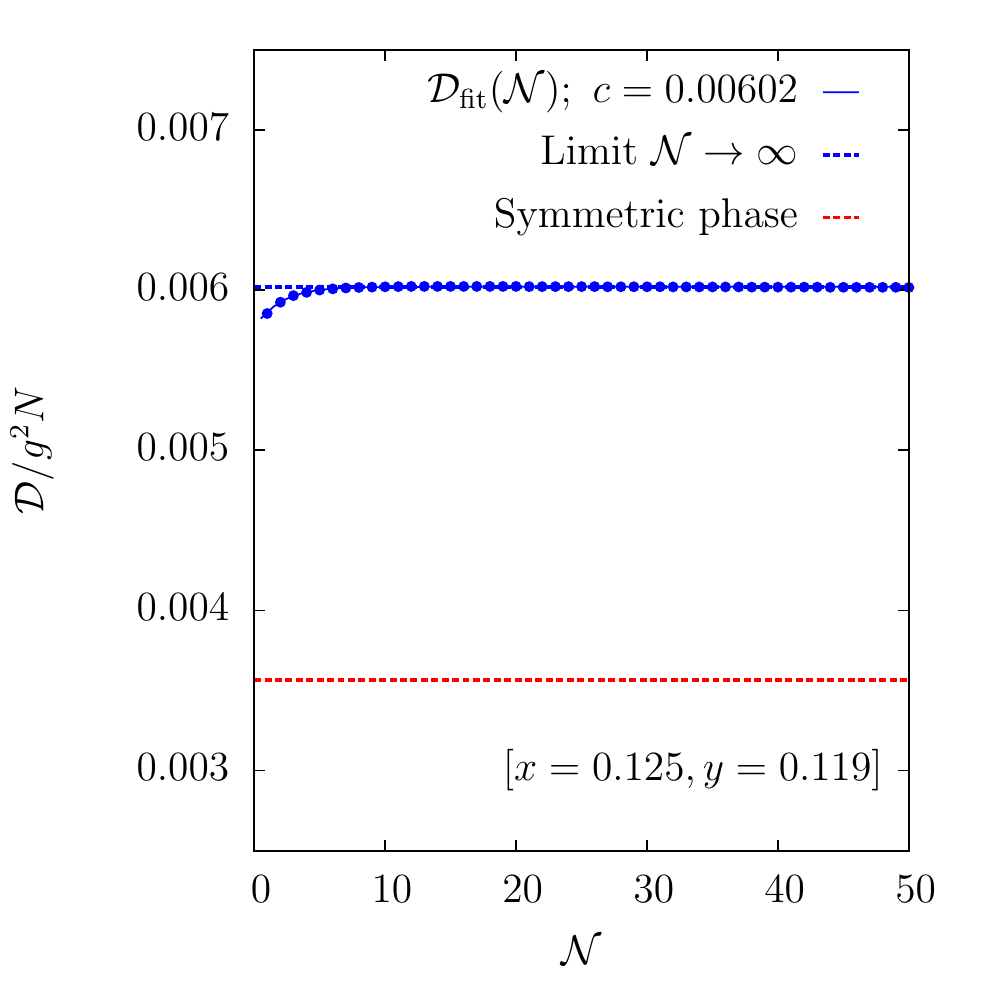}
\includegraphics[scale=0.75]{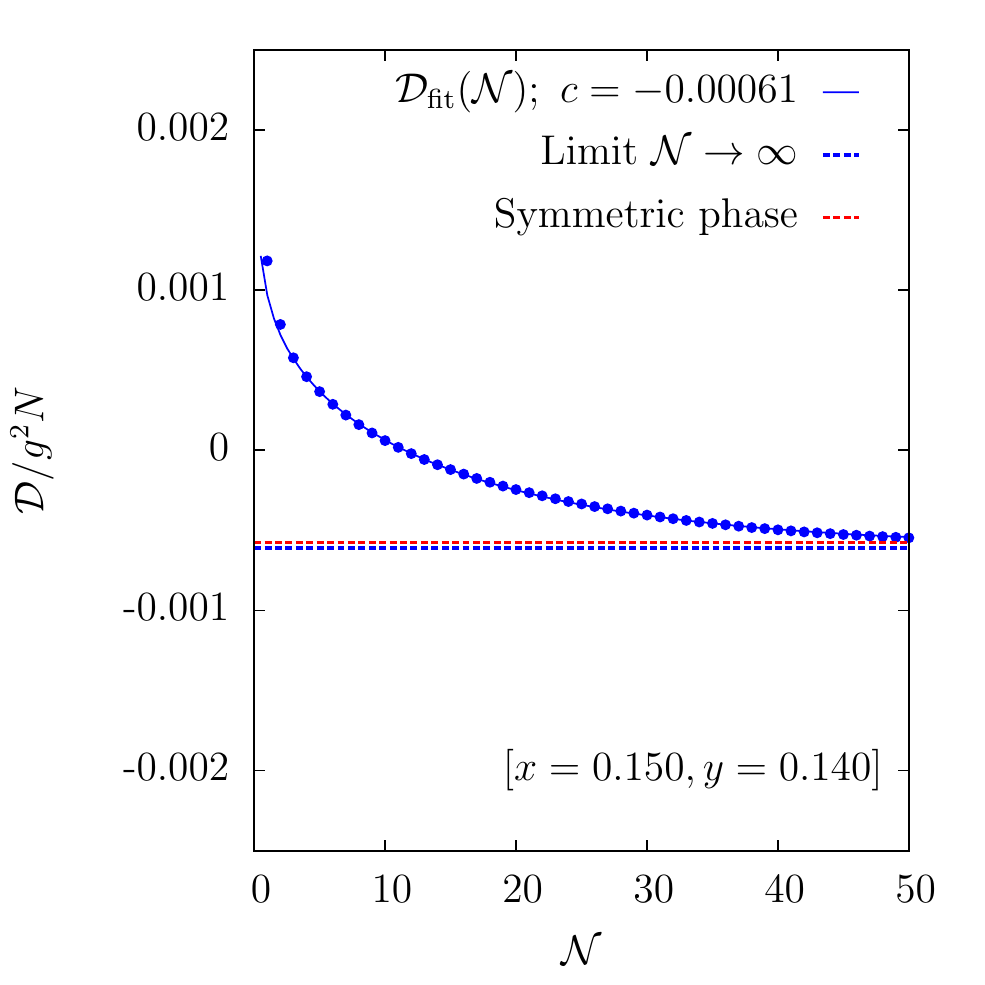}
\caption{\label{fig:VEV_iterations}Convergence of $\Dcond$ (and hence
  $D$) at late times; we show both the cases where $\Dsym$ (red) and $\Dhig$ (blue)
  are distinct (left panel) and equivalent (right panel). At generic
  values of $x$ and $y$, $\Dcond(\mathcal{N})$ will resemble one of
  these two graphs (when it converges).}
\end{figure}

To establish two distinct metastable solutions, it is also important not
to be fooled by slow convergence to an extremal solution.  We illustrate
this idea in Fig.~\ref{fig:VEV_iterations}, which shows algorithm
convergence for two cases.  On the left, we see convergence from two
different starting configurations to two distinct final solutions.  On
the right, we see slow evolution to a single solution.  To distinguish
these cases, it is important to fit the $\mathcal{N}$ dependence of
$\Dcond$ (or some other measure of the solution), to test convergence.
We find that the fit form
\be
\Dcond_{\text{fit}}(\mathcal{N}) = A
 \frac{e^{-\mathcal{N}/\tau_{\mathcal{N}}}}{\mathcal{N}^{\delta}} + c
\ee
gives a good description.

\subsection{Issues of stability}

\label{sec:stability}

Is the extremization of $\Gamma$ a minimization/maximization or a
saddlepoint-seeking procedure?  It is a saddlepoint-seeking procedure,
as can be seen by considering the one-loop value,
\bea
&&\Gamma^{\text{1-loop}}[\GT,\Delta,D] = \nonumber\\
&&\quad \Tr \log D - D D^{(0)-1} + \frac{1}{2}\Tr \log \GT -
\frac{1}{2}\GT {\GT}^{(0)-1} - \Tr \log \Delta + \Delta {\Delta}^{(0)-1}
,\qquad~
\eea
which is solved trivially by $D=D^0$, $G=G^0$, $\Delta = \Delta^0$.
For the sign choice above, this extremum is clearly a maximum for $G$,
but a minimum for $\Delta$, since the ghost enters with the opposite
sign. At least in the ultraviolet this property is not affected by
additional diagrams, since the tree terms dominate in the UV.

This does not cause a problem in practice, since we can alternately
extremize with respect to $G,D$ holding $\Delta$ fixed and with respect
to $\Delta$ holding $G,D$ fixed.  The former involves maximization, the
latter involves minimization.  This procedure shows rapid convergence,
and was used in our previous work \cite{3PI}.  The interpretation of
this saddle behavior is benign; it arises because of the peculiarities
of gauge fixing and the presence of the ``fermionic'' ghost species
which it introduces.

But we are in trouble if the bosonic part of $\Gamma$ at fixed $\Delta$
is unbounded from above and below.
When we move from one to three loops and we consider the possibility of
large infrared contributions in $D$, we find that precisely
this problem arises.  The pure scalar diagrams, omitting group theoretic
factors, are of the form
\bea
-\Gamma^{\text{scalar}}[D] &=& -\Tr \log D + D D^{(0)-1}
+ \lambda \int_{pk} D(p)D(k) \nonumber \\
&&\quad-~
\frac{\lambda^2}{2}\int_{pkq} D(p)D(k)D(q)D(p+k+q)
\label{eq:scalarGamma}
\eea
where we have flipped the overall sign so the one-loop piece
opens upwards. With the three-loop term present, $\Gamma$ is
unbounded from above and below. This unbounded behavior becomes
important whenever $D$ becomes sufficiently large in some narrow
momentum range.  For instance, when a small $p$ range around zero
supports a finite value of the integral $\int_p D(p)$, then the small
$p,k,q$ contribution to the three-loop (basketball) term becomes large, and
it diverges as the phase space region supporting $\int_p D \sim 1$ goes
to zero.

Unfortunately, the expected behavior deep in the Higgs phase is
precisely that $\int_p D(p)$ should receive a finite contribution from a
very narrow momentum range near $p=0$.  Therefore, the extremum we seek
is at best a local maximum as a function of $G,D$; and in particular, we
can expect trouble deep in the Higgs phase.  The origin of this problem
is that, when the field develops large long-distance correlations, the
loopwise expansion of the 2PI functional breaks down.
For instance, at the four-loop level we will encounter
\be
+\lambda^3 \int_{pkql} D(p) D(p+l) D(q) D(q+l)D(k)D(k+l)
\ee
which diverges still more strongly.  The sequence of such divergent
graphs is resummed by including the one-point function in
the procedure, and stripping the square of the one-point function from the
two-point correlator.  However, as we have emphasized, any procedure for
including the one-point function appears to damage the properties which
ensure the possibility of a phase transition endpoint.

Here we work in terms of the two-point function only, which will mean that
we are unable to study cases where the solutions become strongly
Higgs-like.  We anticipate that, when we seek solutions which show
strong Higgs-like behavior, we will instead find runaway behavior in our
extremization algorithm.

Naively it appears that this problem is less severe at small $x$ where
the high-loop Higgs diagrams are suppressed by more explicit powers of
$x$. But the Higgs-phase value of the condensate $\int_p D(p)$ grows as
$1/x^2$, so in fact the problem is more severe, not milder, at small
$x$.  Therefore it will not be possible to make contact with the
perturbative part of the phase diagram.

\section{Analysis and results}

\label{Results}

We will concentrate on the analysis in Landau gauge
(which eliminates the longitudinal gluon propagator), and set $\Nc = 2$
with the scalar field in the fundamental representation. A comparison
with the results in Feynman gauge appears towards the very end, in
Section \ref{sec:compgauge}.

\begin{figure}
\centering
\includegraphics[scale=0.75]{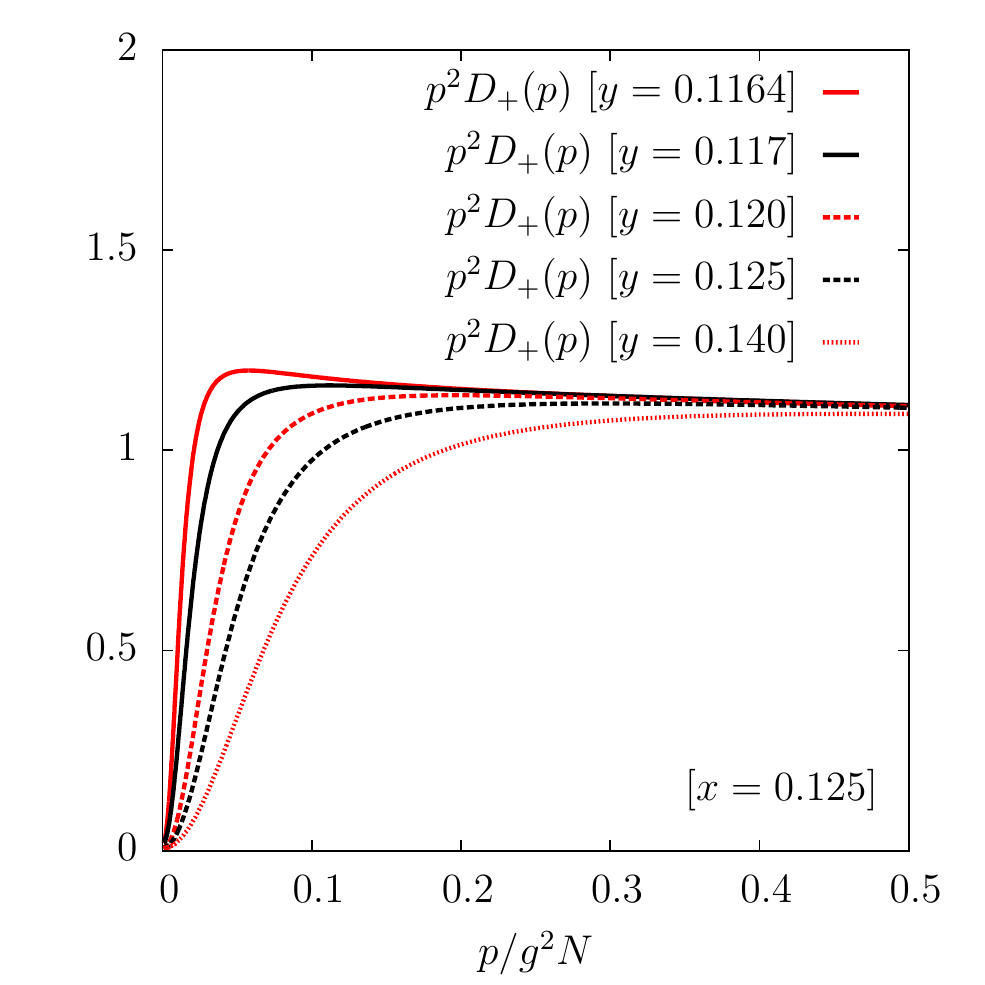}
\includegraphics[scale=0.75]{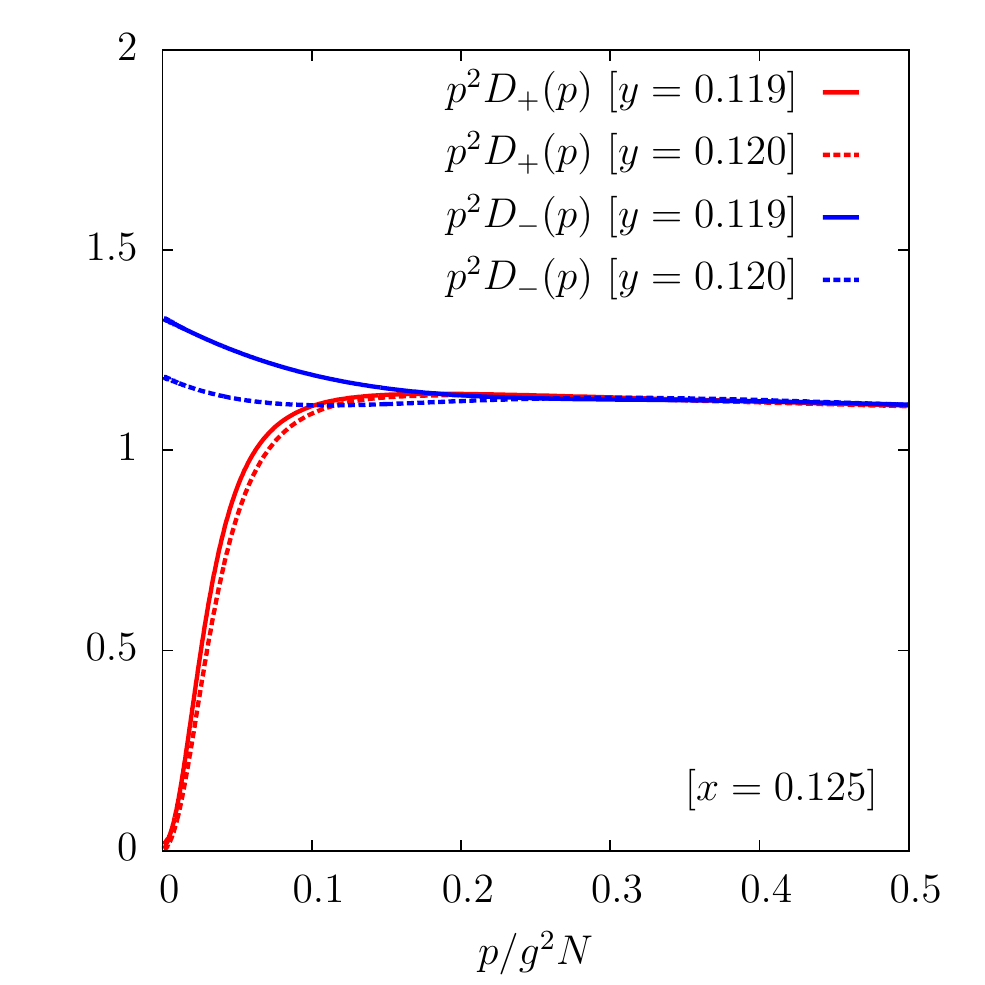}
\caption{\label{fig:G_plots}Evolution of the symmetric phase solution
  for Higgs two-point function with increasing $y$ at fixed $x = 0.125$
  (left), and coexistence of symmetric and Higgs phase solutions
  (right).}
\end{figure}

\begin{figure}
\centering
\includegraphics[scale=0.75]{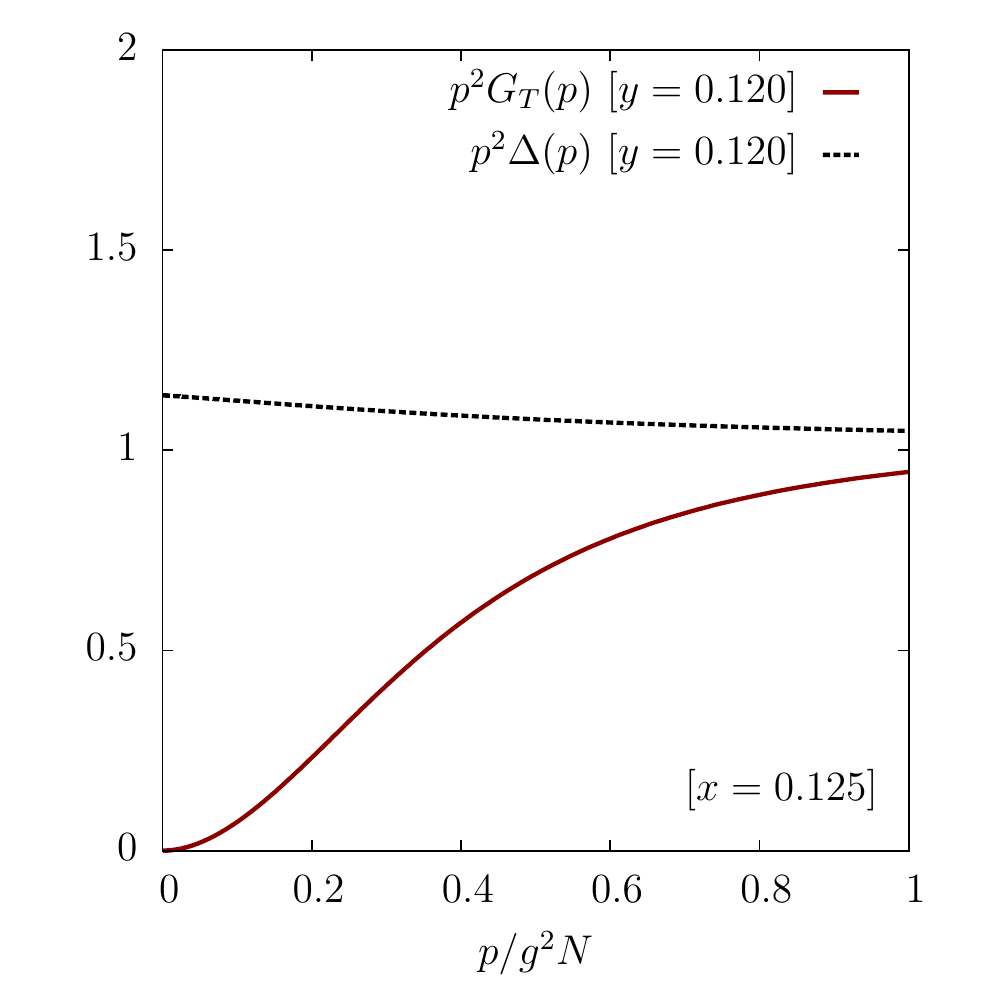}
\caption{\label{fig:D_plots}Transverse gauge field and ghost two-point
  functions (in the symmetric phase), showing nonperturbative massive behavior in $\GT$. The corresponding broken phase solutions are not depicted since for this particular $(x,y)$ they would be nearly indistinguishable on this plot. However, it is worth noting that in general $\Pi_T$ becomes increasingly massive in the Higgs phase relative to the symmetric phase (when these solutions coexist).
}
\end{figure}

Solutions for the Higgs, gauge and ghost propagator are shown in
Fig.~\ref{fig:G_plots} (Higgs) and Fig.~\ref{fig:D_plots}
(gauge/ghost). These plots are generated for the specific value of
$x=0.125$ (and a range of $y$); however, at generic values of $(x,y)$
solutions (when they exist) will take on either of these forms. On the
right panel in Fig.~\ref{fig:G_plots}, we can distinguish between the
peak-like and massive behavior of Higgs ($\Dhig$) and
symmetric ($\Dsym$) phase solutions. Furthermore, at $x=0.125$, we
observe that $\Gamma$ simultaneously admits two solutions over a range
of $y$, which is evidence of metastability. As shown and explained in
Fig.~\ref{fig:D_plots}, both solutions display gauge
correlators with ``massive'' behavior (in the sense that
$\lim_{p\to 0}G_T(p)$ is finite; we are not claiming that the propagator
has a pole at imaginary $p$ and we have not investigated the behavior of
spatial Wilson loops).  In the symmetric phase this is due primarily to
pure-glue loops; in the Higgs phase the mass is larger, due to
additional Higgs-loop contributions.

On the left panel of
Fig.~\ref{fig:G_plots} we see that the symmetric solution $\Dsym$
terminates. This indicates that the symmetric phase has lost its
metastability and become spinodally unstable; so we identify the $y$
value where this occurs as $y_-(x)$.
The Higgs solution also terminates, and the possibility of metastability
ceases to occur, at a larger $y$ value which we interpret as $y_+(x)$.
There is a third special value of $y$, where the Higgs solution becomes
unstable to runaway behavior as described in Subsection
\ref{sec:stability}.  We will label this value $y_{\text{end}}$.  It
does not have a physical interpretation in terms of the phase diagram;
it is simply the point where the solution becomes so Higgs-like that our
three-loop truncation encounters uncontrolled stability issues when we try
to analyze the Higgs branch.

We can map out the region of the $(x,y)$ plane between the
$y_+$ and $y_-$ curves by finding those regions where two (meta)stable
solutions for the propagators exist.  The ``symmetric'' (small-$\Dcond$)
solution is obtained by seeding the gradient solver with a configuration
found at larger $y$, while the ``broken'' solution is found by starting
at a smaller value of $y$ with an initial guess for the scalar
propagator with strong small-$p$ behavior.  The critical value $x_c$ is
the largest $x$ such that metastability is observed.

\begin{figure}
\centering
\includegraphics[scale=0.9]{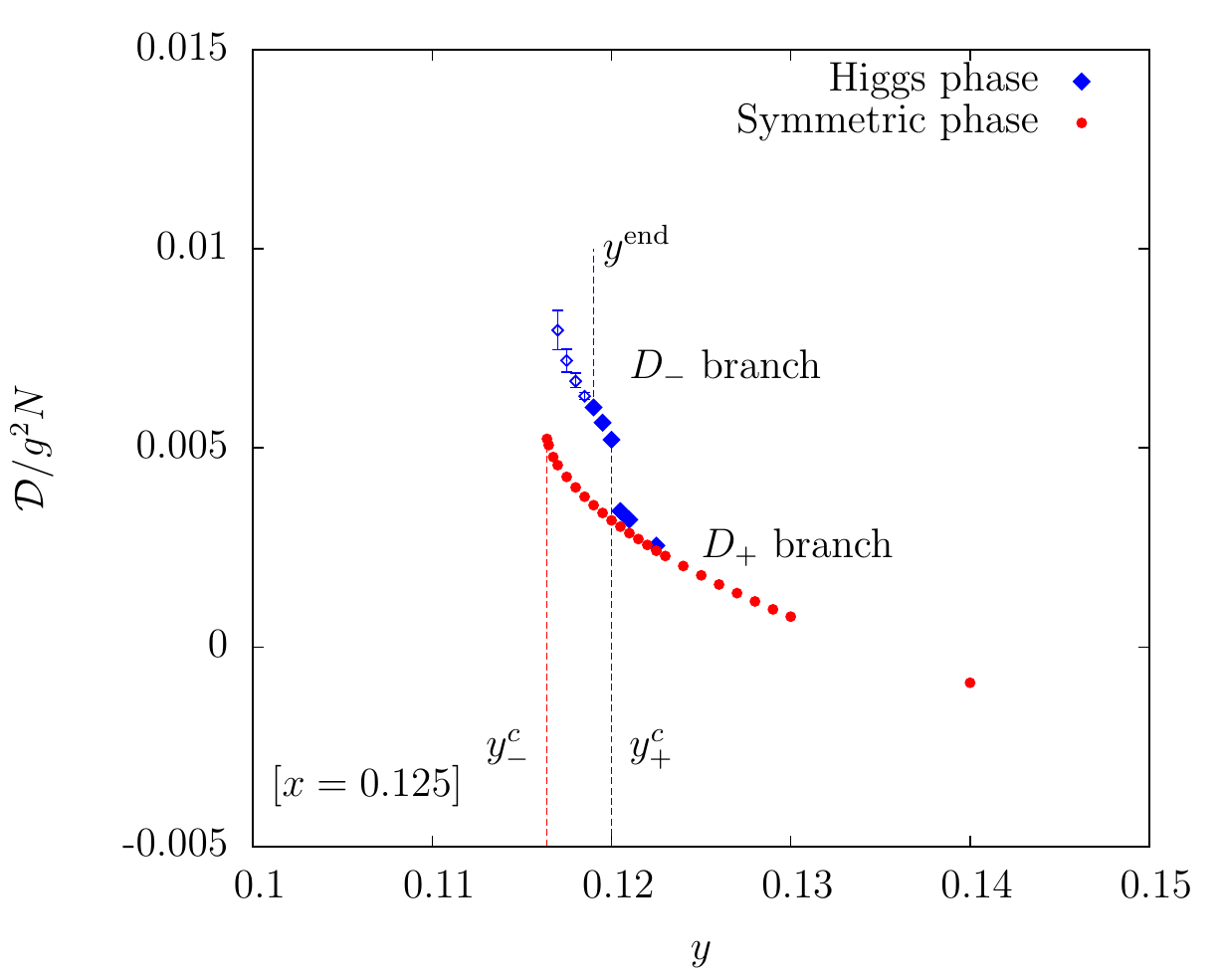}
\includegraphics[scale=0.9]{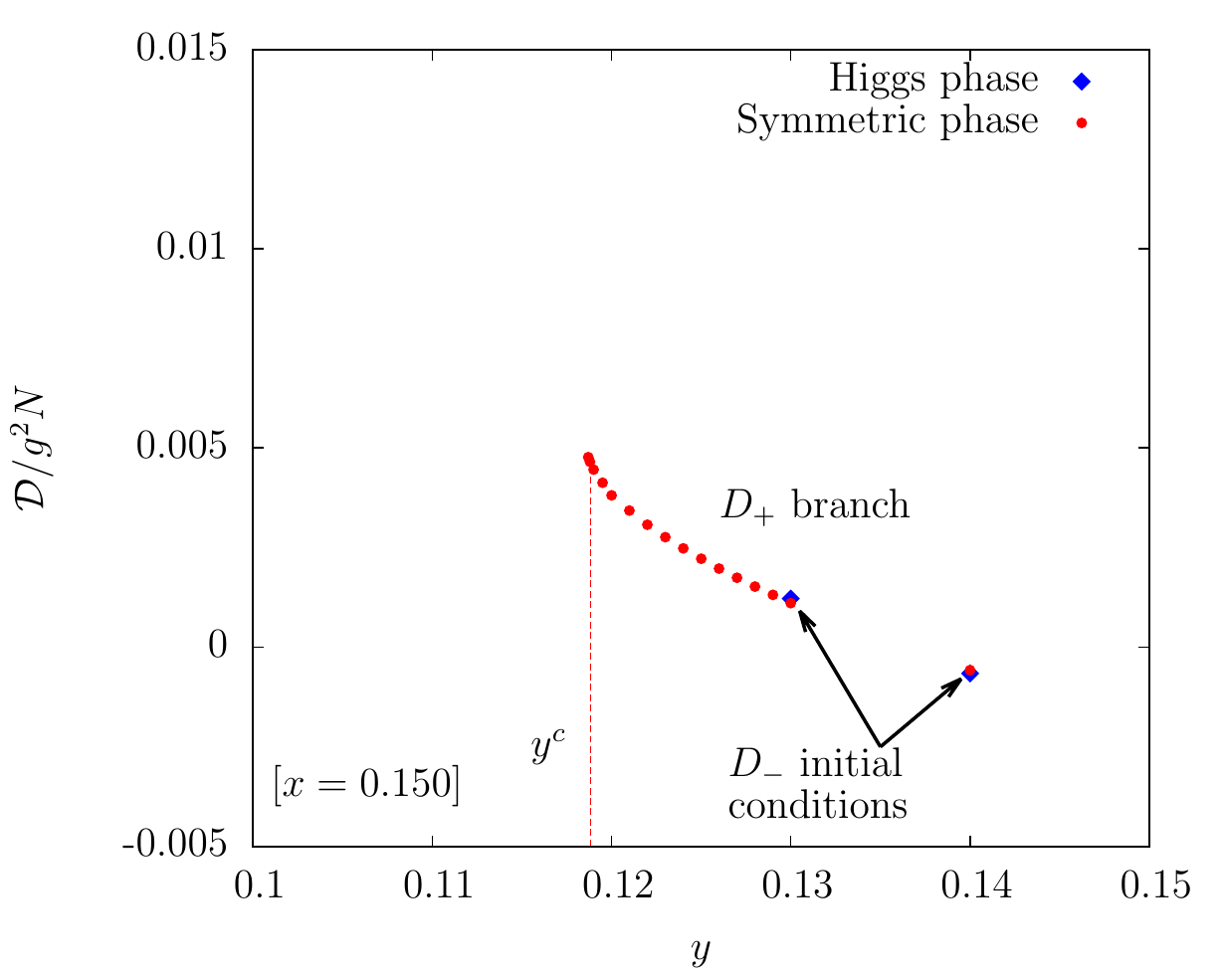}
\caption{\label{fig:VEV_critical} Evolution of $\Dcond$ with $y$ at
  fixed $x$, showing the appearance of stable branch of Higgs phase
  solutions at $x=0.125$. Unstable fixed points are drawn as
  silhouettes. }
\end{figure}

\begin{figure}
\centering
\includegraphics[scale=1]{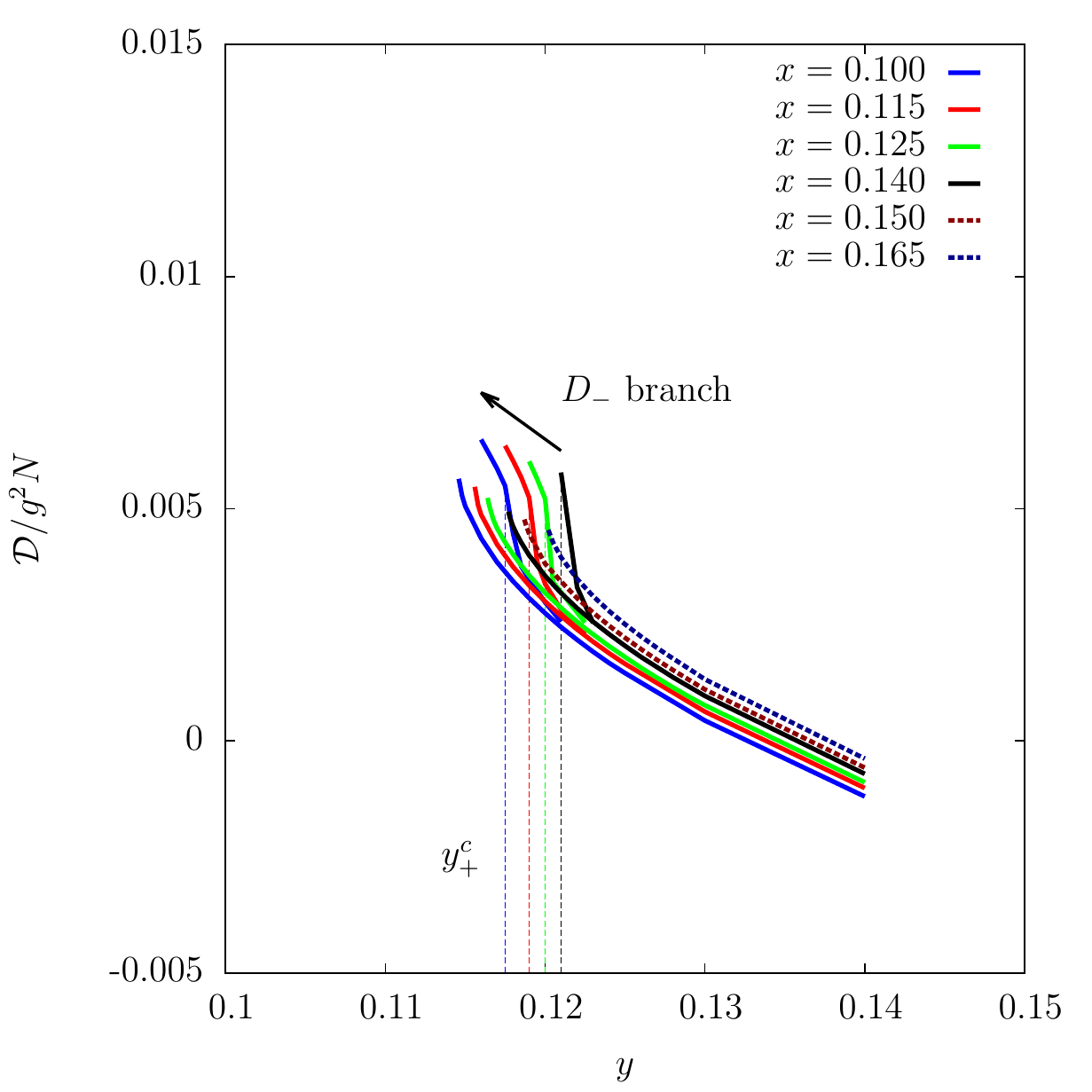}
\caption{\label{fig:VEV_All}Evolution of the stable branches with
  increasing $x$. The $\Dhig$ branch disappears by $x=0.150$, indicating
  the absence of a region of metastability. {}From this we can infer
  $x_c \simeq 0.150$.}
\end{figure}

Plots of $\Dcond(y)$ for $x = 0.125$ and $x = 0.150$ are shown
in Fig.~\ref{fig:VEV_critical}. In both cases we see a branch of
symmetric phase solutions which terminates at $y_-$.
But while $x=0.125$ supports a Higgs branch, $x=0.15$ does not; so $x_c$
must occur between these two values.  To determine where, we carry out a
scan of the phase structure for several values of $x$, as shown in
Fig.~\ref{fig:VEV_All}.  The figure displays two-branch behavior at
$x=0.14$ but not $x=0.15$, so we conclude that
$0.14 < x_c < 0.15$.

\subsection{Comparison with the lattice}

Our original purpose in applying the 2PI formalism to $\SUN$ Higgs
theory was not specifically to determine the phase diagram (which is
already known), but rather, to test the accuracy with which $n$PI
resummation is able to make predictions about the nonperturbative
sector of a nonabelian gauge theory. The $n$PI method relies on
approximating the effective action by its truncation at a finite loop
order, which results in a selective resummation to all orders of a
certain class of topologies. In a gauge theory, this induces
gauge-fixing dependence \cite{Arrizabalaga1,Carrington1}, since at least
perturbatively, one should include all diagrams at every loop
order. This effect could potentially be very mild, but \textit{a priori}
it is not clear that accurate results can be obtained from this method
anywhere on the phase diagram. The only way to test the reliability of
the approximation is to directly compute gauge-invariant observable
quantities.

Here we will attempt a direct comparison between lattice and 2PI
determined values of $x_c,y_{-}(x)$ and $y_+(x)$.
An overview of many of the pertinent results from 3D lattice studies of
SU(2) Higgs theory can be found in \cite{Rummukainen2}, which
incorporates the original studies
\cite{Kajantie2,Kajantie3,Gurtler1,endpoint}.
The most relevant quantity to compare is the location of the critical
endpoint.  We find $(x_c,y_c) \simeq (0.145,0.118)$.  The accepted
nonperturbative lattice value is
$(x_c,y_c) = (0.0983\pm .0015,-.0175\pm .0013)$.  There is qualitative
agreement, but quantitatively the 2PI method has $\sim 50\%$ relative
errors in $x_c$ (establishing relative errors in $y_c$ is harder since
it depended on an arbitrary renormalization point prescription).

We could also try to compare the spacing $y_+ - y_-$ to the lattice, at
a comparable distance below $x_c$.  Unfortunately, the locations of the
upper and lower metastability lines lack a clean nonperturbative
definition.  Technically, at any $(x,y)$ value there is only one possible
phase, and the transition line is where the is an abrupt change in that
phase's properties, such as $\Dcond$.  In practice, for systems near the
transition line there are very long-lived metastable states, and the
transition from the metastable to the stable state involves an extremely
rare and spatially inhomogeneous configuration.  The spatial
inhomogeneity is the reason that our 2PI approach cannot explore such
states, allowing us to explore the supercooled or superheated phases.
The lattice avoids this problem by sampling over all such states,
typically using reweighting to make it more likely to sample the
inhomogeneous states which carry us between metastable and stable
phases.
Nevertheless, Ref.~\cite{Kajantie2} presented a definition of the
metastability limits.  For $x=.0645$ they find
$y_+=.0009$ and $y_-=-.0086$, for a range of $(y_+-y_-)=.0095$.  This is
comparable to the ranges we see in Fig.~\ref{fig:VEV_All} for $x=0.10$.
So there is at least qualitative agreement here.

We are not able to compare the discontinuity between condensates at
$y_c$ as a function of $x_c-x$ to the lattice, because we have not
implemented a procedure to find the $\Gamma$ difference between the two
phases and thereby determine the transition value $y_c$.

\subsection{Comparison between Landau and Feynman gauges}

\label{sec:compgauge}

\begin{figure}
\centering
\includegraphics[scale=1]{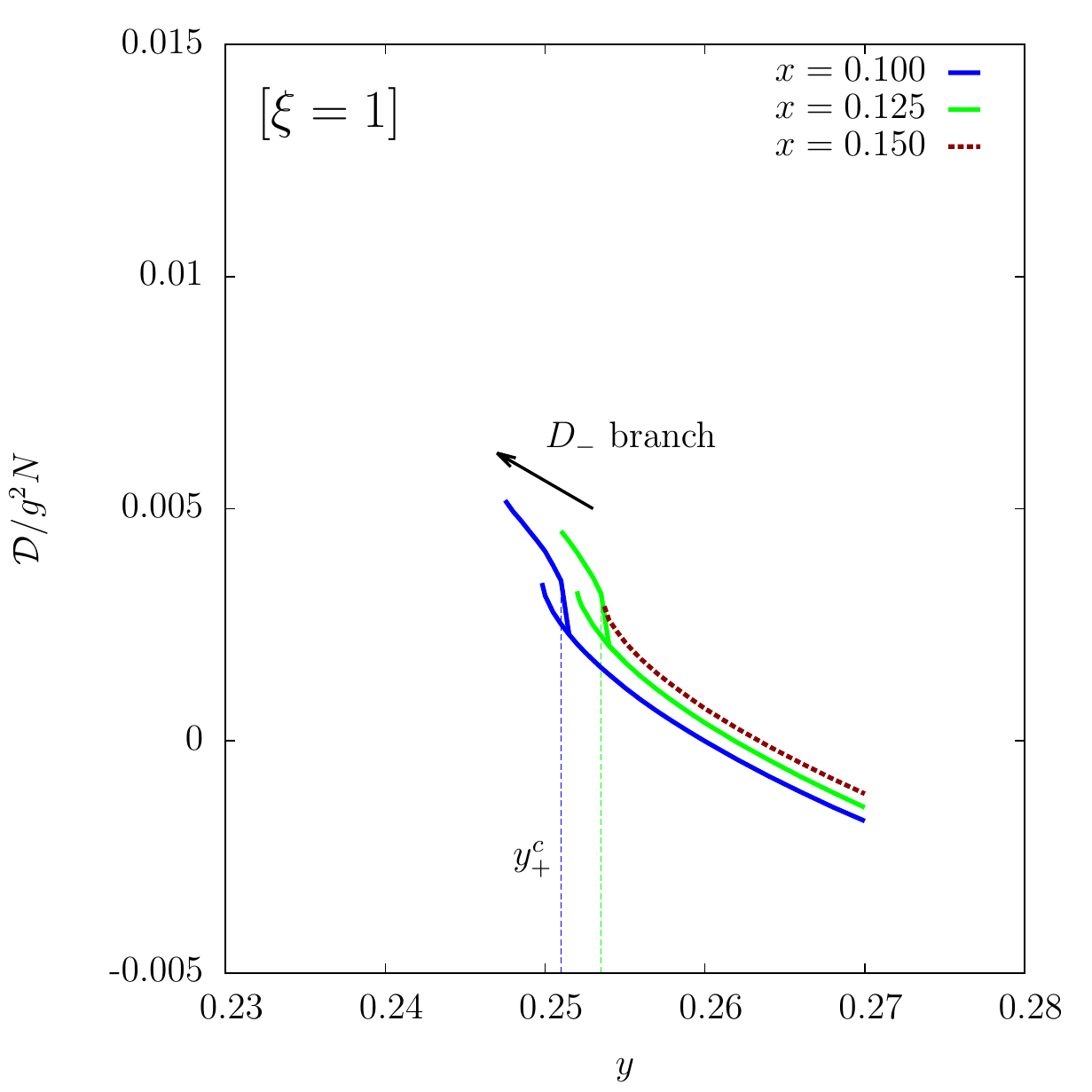}
\caption{\label{fig:VEV_All_XI10}
Feynman gauge analogue of Fig.~\ref{fig:VEV_All}, showing the evolution
of the $\Dsym$ and $\Dhig$ branches with $x$. As in Landau gauge, the
$\Dhig$ branch disappears by $x=0.150$, but the critical range of $y$
has shifted.}
\end{figure}

Up to now, we have argued diagrammatically that critical values of $y$
are expected to exhibit dependence on the gauge parameter
$\xi$. However, since it is difficult to quantify this effect without an
explicit computation, we will now briefly present a comparison between
Landau and Feynman gauges. The results in Feynman gauge are best
summarized by a $\xi = 1$ analogue of Fig.~\ref{fig:VEV_All}, shown in
Fig.~\ref{fig:VEV_All_XI10}.

In setting $\xi = 1$ and resolving the SD equations (following the usual
procedure), we observe that qualitatively very little has
changed. Feynman gauge solutions exhibit similar features to those in
Landau gauge, and once again we observe a disappearance of a stable
Higgs branch somewhere between $x = 0.125$ and $x = 0.150$. This is
consistent with the observation that dependence on $x$ enters primarily
through diagrams without gauge field lines. The biggest change though is
the observed shift in the critical range of $y$, from around
$y \sim 0.120$ to $y \sim 0.250$.
This is interpreted as a contribution to the scalar mass from $G_L$
propagators, which does not fully cancel between diagrams.
For instance, the mass contributions (at vanishing external momentum) of
$G_L$ in the two self-energy corrections
\begin{fmffile}{fmf_Feynmangauge}
\be
\label{eq:gluetadpoleFgauge}
\parbox{20mm}{\begin{fmfgraph}(20,20)
\fmfleft{l1}\fmfright{r1}
\fmfforce{0.25w,0.5h}{v1}\fmfforce{0.75w,0.5h}{v3}
\fmf{dbl_plain}{l1,v1}\fmf{dbl_plain,right=1}{v1,v3}\fmf{dbl_wiggly,right=1}{v3,v1}\fmf{dbl_plain}{v3,r1}
\end{fmfgraph}}
 ~ + \frac{1}{2} ~
 \parbox{20mm}{\begin{fmfgraph}(20,20)
\fmfleft{l1}\fmfright{r1}\fmftop{t1}\fmfforce{0.5w,0.5h}{v1}
\fmf{dbl_plain}{l1,v1}\fmf{phantom,tension=5.0}{t1,v2}
\fmf{dbl_wiggly,left=1,tension=0.4}{v1,v2,v1}
\fmf{dbl_plain}{v1,r1}
\end{fmfgraph}}
\ee
\end{fmffile}
only cancel if the scalar propagator takes the free massless value
$1/p^2$; otherwise the tadpole contribution is larger and leads to a
positive mass contribution which is proportional to the gauge parameter
$\xi$.

\section{Concluding remarks}

We directly solved the three-loop 2PI effective action for 3D SU($\Nc$)
Higgs theory and obtained resummed correlators which correspond to both
the symmetric and Higgs phases of the theory. We found that these
solutions coexist over a region of the phase diagram, indicative of
metastability and a first order phase transition. Subsequently, we have
also observed that there is a point $x$ where the metastability ceases
to be observed, which we identified with the critical end point of the
theory, $x_c$.

Concerning the numerical accuracy of the predictions made in Landau
gauge, the location of the critical end point we inferred differs from
the lattice value with a relative error of $\sim 50\%$.  We also found
that the critical $y_c$ depends surprisingly strongly on the gauge
parameter $\xi$.

The most promising finding regarding the applicability of the $n$PI
formalism to a nonabelian gauge theory is the apparent qualitative
evidence for a critical end point $(x_c,y_c)$ located relatively close
to its known nonperturbative value.  In this sense the 2PI approach has
successfully seen nonperturbative behavior in the phase diagram.
However, the method has shown serious weaknesses as well.  The
quantitative level of agreement with the lattice is not very impressive,
and the strong $\xi$ dependence in $y_c$ is also worrying.  More
urgently, the method has failed completely to resolve the behavior of
the Higgs phase when the scalar condensate is large.  The most
straightforward way to fix this problem, via the introduction of a
scalar one-point function and the use of $R_\xi$ gauge, would introduce
new problems.  As we have argued, the $\xi$ dependence should be
significant where the transition is weak, and the gauge-fixing procedure
may destroy the existence of a critical endpoint and analytic connection
between the two ``phases.''

Thus, the study of SU($\Nc$) Higgs theory has therefore revealed several
limitations to the $n$PI method in the context of a nonabelian gauge
theory. In addition to the described ambiguities in physical
observables, the application of the formalism is \textit{difficult}
numerically, especially if one wishes to consider higher-loop
truncations or higher $n$-particle-irreducibility. If qualitative
predictions can be made at best, then it may be hard to justify the
numerical expense. However, this work does not preclude the possibility
that further refinements may be possible with the goal of obtaining
quantitatively accurate answers to nonperturbative and gauge-invariant
questions. This matter is left open for a future
investigation.

\section*{Acknowledgments}

We would like to thank Meg Carrington and Marcus Tassler for useful
comments.  This work was supported in part by the Natural Science and
Engineering Research Council of Canada (NSERC).

\appendix

\section{\label{FeynRules}Feynman rules for SU($\Nc$) Higgs theory}

The Feynman rules for covariant-gauge perturbative calculations in
SU($\Nc$) Higgs theory are derived from the Lagrangian
\bea
\mathcal{L} &=& \frac{1}{2}\Tr F_{\mu\nu}F^{\mu\nu}
+\frac{1}{2\xi}(\partial^\mu A_\mu^a)^2 + \partial_\mu\bar{c}^a
\partial^\mu c^a - gf_{abc}\partial^\mu \bar{c}^a c^b A_\mu^c\\
 &&+~ (D^\mu \phi)^\dagger (D_\mu\phi) + (m^2 + \delta m^2)
\phi^\dagger \phi + \frac{\lambda}{2}(\phi^\dagger\phi)^2 \,.
\eea
Gauge field, scalar and ghost propagators are denoted by the symbols
$G,D$ and $\Delta$. In Euclidean space at tree level these are
\begin{eqnarray}
G^{(0)}_{\mu\nu}(p) &=& \frac{1}{p^2}\Big (\mathbf{T}_{\mu\nu}(p) +
\xi \mathbf{L}_{\mu\nu}(p) \Big) \\
D^{(0)}(p) &=& \frac{1}{p^2 + m^2} \\
\Delta^{(0)}(p) &=& \frac{1}{p^2}
\end{eqnarray}
where the gauge-field propagator is specified by the transverse and
longitudinal projectors
\bea
\Tproj_{\mu\nu}(p) &=& g_{\mu\nu} - \frac{p_\mu p_\nu}{p^2}  \\
\Lproj_{\mu\nu}(p) &=& \frac{p_\mu p_\nu}{p^2} .
\eea
With all momenta assumed to be flowing outwards, the bare Yang-Mills
vertices are
\bea
gV^{(0)a_1a_2a_3}_{\mu_1\mu_2\mu_3} &=& g F^{a_1a_2a_3}\big(
(p_2-p_3)_{\mu_1}g_{\mu_2\mu_3} + (p_3-p_1)_{\mu_2} g_{\mu_1\mu_3} \nonumber \\
&& {}+
(p_1-p_2)_{\mu_3} g_{\mu_1\mu_2} \big) \qquad~\\
g\mathbb{V}_{\mu_3}^{(0)a_1a_2a_3} &=& g F^{a_1a_2a_3}p_{1\mu_3}\\
\nonumber g^2V^{(0)a_1a_2a_3a_4}_{\mu_1\mu_2\mu_3\mu_4} & = &
g^2\big(F^{a_1a_2s}F^{a_3a_4s}(g_{\mu_1\mu_3}g_{\mu_2\mu_4} - g_{\mu_1\mu_4}g_{\mu_2\mu_3}) \\
\nonumber &&{}+ F^{a_1a_3s}F^{a_4a_2s}(g_{\mu_1\mu_4}g_{\mu_2\mu_3}
                                -g_{\mu_1\mu_2}g_{\mu_3\mu_4})\\
&& {}+ F^{a_1a_4s}F^{a_2a_3s}(g_{\mu_1\mu_2}g_{\mu_3\mu_4}
                             -g_{\mu_1\mu_3}g_{\mu_2\mu_4})\big) .
\eea
where $(a_1,p_1)$ are the color indices and momentum of the outgoing
ghost in $\mathbb{V}$.  The presence of a complex scalar results in
the following additional vertices,
\bea
\label{eq:AppVssg}g \mathcal{V}_{\mu_3}^{(0)a_1a_2a_3} & = &g T^{a_3}_{a_1a_2}(p_1 - p_2)_{\mu_3}\\
\label{eq:AppVssgg} g^2 \mathcal{V}_{\mu_3 \mu_4}^{(0)a_1a_2a_3a_4} & = & -g^2 T^{\{a_3}_{a_1 s}T^{a_4\}}_{s a_2}g_{\mu_3\mu_4}\\
\label{eq:AppVssss} \lambda \mathcal{V}^{(0)a_1a_2a_3a_4} &=& -\lambda (\delta_{a_1a_2}\delta_{a_3a_4}+\delta_{a_1a_4}\delta_{a_2a_3}) .
\eea 
where the outgoing scalar(s) are indexed by $(a_1,p_1)$
(\Eq{eq:AppVssg} and \Eq{eq:AppVssgg}) and $a_1,a_3$
(\Eq{eq:AppVssss}).

\section{\label{SelfEnergy}Self-energies computed in dimensional regularization}

In regularizing the 2PI effective action, one makes use of one- and
two-loop self-energy corrections computed in perturbation theory. In
pure Yang-Mills, all one- and two-loop integrals are massless from the
onset. However, the inclusion of a Higgs field now in principle adds
massive propagators to many of the diagrams. But, since we really only
need to know the UV limit of these diagrams, it actually suffices to
compute them with a massless scalar field.

In this Appendix, though some results are valid for arbitrary
$\Dim$, $\epsilon$ should be treated as a small parameter,
\textsl{i.e.}, it is assumed that we are working at or near 3
dimensions, $\Dim = \text{D}_0 + 2\epsilon$, with $\text{D}_0 = 3$.
Finally, since the Higgs mass renormalizes at the two-loop level
in three dimensions, it is useful to define the $\MSbar$ scale
$\mubar = \mu^2 e^{\gamma}/4\pi$. The master one-loop topology is
\begin{fmffile}{fmf_oneloop}
\be
\parbox{30mm}{\begin{fmfgraph*}(30,20)
\fmfleft{l1}\fmfright{r1}
\fmfforce{0.25w,0.5h}{v1}\fmfforce{0.75w,0.5h}{v3}
\fmf{fermion,label=$p$}{l1,v1}
\fmf{fermion,left=1,label=$p_1$}{v1,v3}
\fmf{fermion,left=1,label=$p_2$}{v3,v1}
\fmf{plain}{v3,r1}
\end{fmfgraph*}} ~ =  J^{(\Dim)}_1(n_1,m_1;n_2,m_2)
\qquad 
\left \{ \begin{array}{ccc} 
p_1 &=& q \\
p_2 &=& q - p
\end{array} \right .\\
\ee\vspace{5mm}
with
\be
 J^{(\Dim)}_1(n_1,m_1;n_2,m_2) = \Big (\frac{1}{\mubar}\Big
 )^{\frac{\Dim-\text{D}_0}{2}}\int \frac{d^{\Dim} q}{(2\pi)^{\Dim}}
 \frac{1}{\big (q^2 + m_1^2\big )^{n_1}\big ((q-p)^2 + m_2^2\big
   )^{n_2}} .
\ee

\subsection{One-loop gluon self-energy}

The presence of a scalar field adds two additional diagrams to the
one-loop gluon self-energy relative to the the pure Yang-Mills
expression,
\bea
\Pi^{(1,\ep)}_{m^2;\mu\nu} &=&  \frac{1}{2}~
\parbox{20mm}{\begin{fmfgraph}(20,20)
\fmfleft{l1}\fmfright{r1}
\fmfforce{0.25w,0.5h}{v1}\fmfforce{0.75w,0.5h}{v3}
\fmf{photon}{l1,v1}\fmf{photon,left=1}{v1,v3,v1}\fmf{photon}{r1,v3}
\end{fmfgraph}}
~ + \frac{1}{2}~
\parbox{20mm}{\begin{fmfgraph}(20,20)
\fmfleft{l1}\fmfright{r1}\fmftop{t1}\fmfforce{0.5w,0.5h}{v1}
\fmf{photon}{l1,v1}\fmf{phantom,tension=5.0}{t1,v2}
\fmf{photon,left=1,tension=0.4}{v1,v2,v1}
\fmf{photon}{r1,v1}
\end{fmfgraph}}
~ - ~
\parbox{20mm}{\begin{fmfgraph}(20,20)
\fmfleft{l1}\fmfright{r1}
\fmfforce{0.25w,0.5h}{v1}\fmfforce{0.75w,0.5h}{v3}
\fmf{photon}{l1,v1}\fmf{dots_arrow,left=1}{v1,v3,v1}
\fmf{photon}{r1,v3}
\end{fmfgraph}}\nonumber\\
&& +~
\parbox{20mm}{\begin{fmfgraph}(20,20)
\fmfleft{l1}\fmfright{r1}
\fmfforce{0.25w,0.5h}{v1}\fmfforce{0.75w,0.5h}{v3}
\fmf{photon}{l1,v1}\fmf{fermion,left=1}{v1,v3,v1}\fmf{photon}{r1,v3}
\end{fmfgraph}}
~ + ~
\parbox{20mm}{\begin{fmfgraph}(20,20)
\fmfleft{l1}\fmfright{r1}\fmftop{t1}\fmfforce{0.5w,0.5h}{v1}
\fmf{photon}{l1,v1}\fmf{phantom,tension=5.0}{t1,v2}
\fmf{fermion,left=1,tension=0.4}{v1,v2,v1}
\fmf{photon}{r1,v1}
\end{fmfgraph}} ~ .
\eea
The result is strictly transverse; we will separate the Yang-Mills and
Higgs contributions as follows,
\begin{eqnarray}
\Pi^{(1,0)}_{m^2;\mu\nu}&=& g^2 p \left ( \pi^{(1,0)}_{\YM} +\pi^{(1,0)}_{m^2} \right )\Tproj_{\mu\nu}\\
\Pi^{(1,\epsilon)}_{0;\mu\nu}&=& g^2 \left ( \frac{p^{1+2\ep}}{\mu^{2\epsilon}}\right ) \left ( \pi^{(1,\ep)}_{\YM} +\pi^{(1,\ep)}_{0} \right )\Tproj_{\mu\nu}.
\end{eqnarray}
The terms which appear in the limit $\Dim\rightarrow 3$ are
\bea
\pi^{(1,0)}_{\YM} &=& \frac{C_A}{64}(\xi^2+2\xi+11)\\
\pi^{(1,0)}_{m^2} &=& - \frac{T_R}{16\pi}\left (-\frac{4m}{p} + \frac{4m^2 + p^2}{p^2} \left ( \pi -2\arctan \frac{2m}{p} \right ) \right )
\eea
and it is also useful to take the $m \rightarrow 0$ limit and keep
terms $\bigO(\ep)$,
\bea
\pi^{(1,\ep)}_{\YM} &=&\frac{C_A}{64}\left ( (\xi^2+2\xi+11)(1-2\epsilon\log 2)+
\ep(12-12\xi-2\xi^2) \right) \label{eq:ApppioneYM}\\
\pi^{(1,\ep)}_{0} &=& -\frac{T_R}{16}\left ( 1 - 2 \ep \log 2 -\ep\right ) .
\eea

\subsection{One-loop Higgs self-energy}

The calculation of the one-loop correction of the Higgs self-energy
proceeds forward in much the same manner,
\bea
\Pi^{(1,\ep)}_{\phi;m^2} &=& ~
\parbox{20mm}{\begin{fmfgraph}(20,20)
\fmfleft{l1}\fmfright{r1}
\fmfforce{0.25w,0.5h}{v1}\fmfforce{0.75w,0.5h}{v3}
\fmf{fermion}{l1,v1}\fmf{fermion,right=1}{v1,v3}\fmf{photon,right=1}{v3,v1}\fmf{fermion}{v3,r1}
\end{fmfgraph}}
~ + 2~
\parbox{20mm}{\begin{fmfgraph}(20,20)
\fmfleft{l1}\fmfright{r1}\fmftop{t1}\fmfforce{0.5w,0.5h}{v1}
\fmf{fermion}{l1,v1}\fmf{phantom,tension=5.0}{t1,v2}
\fmf{fermion,left=1,tension=0.4}{v1,v2,v1}
\fmf{fermion}{v1,r1}
\end{fmfgraph}}
~ + \frac{1}{2} ~
\parbox{20mm}{\begin{fmfgraph}(20,20)
\fmfleft{l1}\fmfright{r1}\fmftop{t1}\fmfforce{0.5w,0.5h}{v1}
\fmf{fermion}{l1,v1}\fmf{phantom,tension=5.0}{t1,v2}
\fmf{photon,left=1,tension=0.4}{v1,v2,v1}
\fmf{fermion}{v1,r1}
\end{fmfgraph}}
\eea
with
\bea
\Pi^{(1,0)}_{\phi;m^2} &=& g^2 p ~\pi^{(1,0)}_{\phi;m^2}\\
\Pi^{(1,\ep)}_{\phi;0} &=& g^2 \left ( \frac{p^{1+2\ep}}{\mu^{2\ep}} \right ) \pi^{(1,\ep)}_{\phi;0} .
\eea
For the $\Dim \rightarrow 3$ and massless limits we have
\bea
\pi^{(1,0)}_{\phi;m^2}  &=& \frac{(1 + d_R) x}{2\pi} \frac{m}{p} + \frac{C_R}{4\pi} \left ( (2-\xi)\frac{m}{p} + \frac{2(p^2-m^2)}{p^2}\arctan \frac{p}{m} \right ) \\
\pi^{(1,\ep)}_{\phi;0}  &=& \frac{C_R}{4}\left ( 1 - 2\ep\log 2 + \ep(1-\xi)  \right ) .
\eea
in terms of the dimensionless quartic coupling $x =  \lambda / g^2$.

\subsection{One-loop ghost self-energy}

The one-loop ghost self-energy is constructed out of a a single diagram,
\begin{equation}
 \Sigma^{(1,\epsilon)} = ~
\parbox{20mm}{\begin{fmfgraph}(20,20)
\fmfleft{l1}\fmfright{r1}
\fmfforce{0.25w,0.5h}{v1}\fmfforce{0.75w,0.5h}{v3}
\fmf{dots}{l1,v1}
\fmf{photon,left=1}{v1,v3}\fmf{dots_arrow,right=1}{v1,v3}
\fmf{dots}{r1,v3}
\end{fmfgraph}}
\end{equation}
for which in $\Dim=3+2\ep$, $\xi$ dependence only appears at $\bigO(\ep)$,
\begin{equation}
\Sigma^{(1,\epsilon)} = g^2 \left ( \frac{p^{1+2\ep}}{\mu^{2\ep}}\right ) \sigma^{(1,\ep)} = g^2 \left ( \frac{p^{1+2\ep}}{\mu^{2\ep}}\right ) \frac{C_A}{16} \left ( 1 - 2\ep \log 2 + \ep(1-\xi) \right ) .
\end{equation}
\end{fmffile}

\subsection{Two-loop topologies}\begin{fmffile}{fmf_twoloop}

The massless two-loop master topology is
\be
\parbox{40mm}{\begin{fmfgraph*}(40,20)
\fmfleft{l1}\fmfright{r1}\fmfforce{0.25w,0.5h}{v1}\fmfforce{0.75w,0.5h}{v2}
\fmfforce{0.5w,1h}{vt}\fmfforce{0.5w,0h}{vb}
\fmf{fermion,label=$p$}{l1,v1}
\fmf{fermion,left=0.45,label=$p_1$}{v1,vt}
\fmf{fermion,left=0.45,label=$p_2$}{vt,v2}
\fmf{fermion,left=0.45,label=$p_3$}{v2,vb}
\fmf{fermion,left=0.45,label=$p_4$}{vb,v1}
\fmf{fermion,label=$p_5$}{vt,vb}
\fmf{plain}{v2,r1}
\end{fmfgraph*}}~= J^{(\Dim)}_2(n_1,n_2,n_3,n_4,n_5)
\qquad 
\left \{ \begin{array}{ccc} 
p_1 &=& q_1 \\
p_2 &=& q_2 \\
p_3 &=& q_1 - p\\
p_4 &=& q_2 - p\\
p_5 &=& q_1 - q_2
\end{array} \right .\\
\ee
\bea
&&J^{(\Dim)}_2 (n_1,n_2,n_3,n_4,n_5) = \Big ( \frac{1}{\mubar}\Big
)^{\Dim-\text{D}_0}\int \frac{d^{\Dim} q_1}{(2\pi)^{\Dim}}
\frac{d^{\Dim} q_2}{(2\pi)^{\Dim}} \nonumber\\
&& \qquad\qquad\frac{1}{\big (q_1^2 \big )^{n_1}
\big (q_2^2\big )^{n_2}
\big ((q_1-p)^2\big )^{n_3}
\big ((q_2-p)^2 \big )^{n_4}
\big ((q_1-q_2)^2 \big )^{n_5}} .\qquad\quad
\eea
The remaining two topologies are related to $J^{(\Dim)}_2$ by
shrinking one or more of the propagators to a point, for instance
\bea
J^{(\Dim)}_2(n_1,n_2,n_3,n_4,0) &=& J^{(\Dim)}_1(n_1,0;n_3,0)J^{(\Dim)}_1(n_2,0;n_4,0)\\
J^{(\Dim)}_2(n_1,0,0,n_2,n_3) &=& J^{(\Dim)}_2 (n_1,n_2,n_3) \\
J^{(\Dim)}_2(n_1,n_2,n_3,0,n_4) &=& J^{(\Dim)}_2(n_1,n_2,n_3,n_4)
\eea
where the number of propagators should be inferred from the
arguments. In computing the two-loop self-energies we encounter UV
divergences arising from the integrals
\bea
J^{(\Dim)}_1(n_1,0;n_2,m) &=&
(m^2)^{\Dim/2-\alpha-\beta}\frac{\Gamma(\Dim/2-n_1)\Gamma(n_1+n_2-\Dim/2)}{(\mubar)^{\frac{\Dim-\Dim_0}{2}}(4\pi)^{\Dim/2}\Gamma(\Dim/2)\Gamma(n_2)}
\nonumber\\
&& \times~ {_2F_1}\Big(n_1,n_1+n_2-\frac{\Dim}{2};\frac{\Dim}{2}
\Big\vert-\frac{p^2}{m^2}\Big)~~~\\
J^{(\Dim)}_2(n_1,n_2,n_3) &=&
\frac{\Gamma(\Dim/2-n_1)\Gamma(\Dim/2-n_2)\Gamma(\Dim/2-n_3)}{(\mubar)^{\Dim-\Dim_0}(4\pi)^{\Dim}\Gamma(n_1)\Gamma(n_2)\Gamma(n_3)}
\nonumber\\
&& \times~ \frac{\Gamma(n_1 + n_2 + n_3-\Dim)}{\Gamma (3\Dim/2-n_1 - n_2 - n_3 )}(p^2)^{\Dim-n_1 - n_2 - n_3} .~~~~~~~
\eea
The massive one-loop scalar integral is needed since recursively
one-loop diagrams (i.e. the one-loop diagrams with a self-energy
insertion in one of the propagators) are IR divergent when they are
massless.

\subsection{Two-loop gluon self-energy}

At two loops, a number of additional diagrams are present,
\bea
\nonumber \pi^{(\UV2,\epsilon)}_{\YM;\mu\nu} &\propto& \frac{1}{6}~
\parbox{20mm}{\begin{fmfgraph}(20,20)
\fmfleft{l1}\fmfright{r1}\fmfforce{0.25w,0.5h}{v1}\fmfforce{0.75w,0.5h}{v2}
\fmf{photon}{l1,v1}\fmf{photon,left=1}{v1,v2,v1}
\fmf{photon}{v1,v2}
\fmf{photon}{r1,v2}
\end{fmfgraph}}
~ + \frac{1}{2}~
\parbox{20mm}{\begin{fmfgraph}(20,20)
\fmfleft{l1}\fmfright{r1}\fmfforce{0.25w,0.5h}{v1}\fmfforce{0.75w,0.5h}{v2}
\fmfforce{0.5w,0.75h}{vt}\fmfforce{0.5w,0.25h}{vb}
\fmf{photon}{l1,v1}
\fmf{photon,left=0.5}{v1,vt,v2,vb,v1}
\fmf{photon}{r1,v2}\fmf{photon}{vt,vb}
\end{fmfgraph}}
~ + ~
\parbox{20mm}{\begin{fmfgraph}(20,20)
\fmfleft{l1}\fmfright{r1}\fmfforce{0.25w,0.5h}{v1}\fmfforce{0.75w,0.5h}{v2}
\fmfforce{0.5w,0.75h}{vt}\fmfforce{0.5w,0.25h}{vb}
\fmf{photon}{l1,v1}
\fmf{photon,left=0.5}{v1,vt,v2,vb,v1}
\fmf{photon}{r1,v2}
\fmf{photon,right=0.5}{vt,v2}
\end{fmfgraph}}
~ + \frac{1}{4}~
\parbox{20mm}{\begin{fmfgraph}(20,20)
\fmfleft{l1}\fmfright{r1}\fmfforce{0.25w,0.5h}{v1}\fmfforce{0.75w,0.5h}{v3}
\fmfforce{0.5w,0.5h}{v2}
\fmf{photon}{l1,v1}
\fmf{photon,left=1}{v1,v2,v1}
\fmf{photon}{r1,v3}
\fmf{photon,left=1}{v3,v2,v3}
\end{fmfgraph}}
\\
&& - ~
\parbox{20mm}{\begin{fmfgraph}(20,20)
\fmfleft{l1}\fmfright{r1}\fmfforce{0.25w,0.5h}{v1}\fmfforce{0.75w,0.5h}{v2}
\fmfforce{0.5w,0.75h}{vt}\fmfforce{0.5w,0.25h}{vb}
\fmf{photon}{l1,v1}
\fmf{dots_arrow,left=0.5}{v1,vt}
\fmf{dots,left=0.5}{vt,v2}
\fmf{dots_arrow,left=0.5}{v2,vb}
\fmf{dots,left=0.5}{vb,v1}
\fmf{photon}{r1,v2}\fmf{photon}{vt,vb}
\end{fmfgraph}}
~ - 2 ~
\parbox{20mm}{\begin{fmfgraph}(20,20)
\fmfleft{l1}\fmfright{r1}\fmfforce{0.25w,0.5h}{v1}\fmfforce{0.75w,0.5h}{v2}
\fmfforce{0.5w,0.75h}{vt}\fmfforce{0.5w,0.25h}{vb}
\fmf{photon}{l1,v1}
\fmf{dots,left=0.5}{v1,vt}
\fmf{dots,left=0.5}{vb,v1}
\fmf{photon,left=0.5}{vt,v2,vb}
\fmf{photon}{r1,v2}\fmf{dots_arrow}{vt,vb}
\end{fmfgraph}}
~ - 2~
\parbox{20mm}{\begin{fmfgraph}(20,20)
\fmfleft{l1}\fmfright{r1}\fmfforce{0.25w,0.5h}{v1}\fmfforce{0.75w,0.5h}{v2}
\fmftop{vt}\fmfforce{0.5w,0.25h}{vb}
\fmf{photon}{l1,v1}
\fmf{photon}{r1,v2}
\fmf{dots,left=0.5,tension=3}{v1,v3}\fmf{dots,left=0.5,tension=3}{v4,v2}
\fmf{photon,left=1,tension=0.4}{v3,v4}\fmf{dots,left=1,tension=0.4}{v4,v3}
\fmf{dots_arrow,left=1}{v2,v1}
\fmf{phantom,tension=3}{v3,vt,v4}
\end{fmfgraph}} ~~~
\\
 \pi^{(\IR2,\epsilon)}_{\YM;0;\mu\nu} &\propto& 
\parbox{20mm}{\begin{fmfgraph*}(20,20)
\fmfleft{l1}\fmfright{r1}
\fmfforce{0.25w,0.5h}{v1}\fmfforce{0.75w,0.5h}{v3}
\fmfforce{0.5w,0.25h}{v2}\fmfforce{0.5w,0.75h}{v4}
\fmf{photon}{l1,v1}\fmf{photon,left=0.5,tension=0.4}{v1,v4,v3,v2,v1}
\fmf{photon}{r1,v3}
\fmfIR{v4}
\end{fmfgraph*}}
~+ \frac{1}{2}~
\parbox{20mm}{\begin{fmfgraph*}(20,20)
\fmfleft{l1}\fmfright{r1}\fmftop{t1}\fmfforce{0.5w,0.5h}{v1}
\fmfforce{0.5w,0.5h}{v3}
\fmf{photon}{l1,v1}\fmf{phantom,tension=5.0}{t1,v2}
\fmf{photon,left=1,tension=0.4}{v1,v2,v1}
\fmf{photon}{r1,v1}
\fmffreeze
\fmfIR{v2}
\end{fmfgraph*}}\\
\pi^{(\UV2,\epsilon)}_{0;\mu\nu} &\propto& ~
\parbox{20mm}{\begin{fmfgraph}(20,20)
\fmfleft{l1}\fmfright{r1}\fmfforce{0.25w,0.5h}{v1}\fmfforce{0.75w,0.5h}{v2}
\fmf{photon}{l1,v1}\fmf{fermion,left=1}{v1,v2,v1}
\fmf{photon}{v1,v2}
\fmf{photon}{r1,v2}
\end{fmfgraph}}
~ + ~
\parbox{20mm}{\begin{fmfgraph}(20,20)
\fmfleft{l1}\fmfright{r1}\fmfforce{0.25w,0.5h}{v1}\fmfforce{0.75w,0.5h}{v2}
\fmfforce{0.5w,0.75h}{vt}\fmfforce{0.5w,0.25h}{vb}
\fmf{photon}{l1,v1}
\fmf{fermion,left=0.5}{v1,vt}
\fmf{plain,left=0.5}{vt,v2}
\fmf{fermion,left=0.5}{v2,vb}
\fmf{plain,left=0.5}{vb,v1}
\fmf{photon}{r1,v2}\fmf{photon}{vt,vb}
\end{fmfgraph}}
~ + 2~
\parbox{20mm}{\begin{fmfgraph}(20,20)
\fmfleft{l1}\fmfright{r1}\fmfforce{0.25w,0.5h}{v1}\fmfforce{0.75w,0.5h}{v2}
\fmfforce{0.5w,0.75h}{vt}\fmfforce{0.5w,0.25h}{vb}
\fmf{photon}{l1,v1}
\fmf{plain,left=0.5}{v1,vt}
\fmf{plain,left=0.5}{vb,v1}
\fmf{photon,left=0.5}{vt,v2,vb}
\fmf{photon}{r1,v2}\fmf{fermion}{vt,vb}
\end{fmfgraph}}
~ + 4 ~
\parbox{20mm}{\begin{fmfgraph}(20,20)
\fmfleft{l1}\fmfright{r1}\fmfforce{0.25w,0.5h}{v1}\fmfforce{0.75w,0.5h}{v2}
\fmfforce{0.5w,0.75h}{vt}\fmfforce{0.5w,0.25h}{vb}
\fmf{photon}{l1,v1}
\fmf{fermion,left=0.5}{v1,vt}
\fmf{plain,left=0.5}{vt,v2}
\fmf{fermion,left=0.5}{v2,vb}
\fmf{plain,left=0.5}{vb,v1}
\fmf{photon}{r1,v2}
\fmf{photon,right=0.5}{vt,v2}
\end{fmfgraph}} \\
 \pi^{(\IR2,\epsilon)}_{0;\mu\nu} &\propto& 
\parbox{20mm}{\begin{fmfgraph*}(20,20)
\fmfleft{l1}\fmfright{r1}
\fmfforce{0.25w,0.5h}{v1}\fmfforce{0.75w,0.5h}{v3}
\fmfforce{0.5w,0.25h}{v2}\fmfforce{0.5w,0.75h}{v4}
\fmf{photon}{l1,v1}\fmf{plain,left=0.5}{v1,v4,v3}
\fmf{fermion,left=1}{v3,v1}
\fmf{photon}{r1,v3}
\fmfIR{v4}
\end{fmfgraph*}}
~ + ~
\parbox{20mm}{\begin{fmfgraph*}(20,20)
\fmfleft{l1}\fmfright{r1}\fmftop{t1}\fmfforce{0.5w,0.5h}{v1}
\fmfforce{0.5w,0.5h}{v3}
\fmf{photon}{l1,v1}\fmf{phantom,tension=5.0}{t1,v2}
\fmf{fermion,left=1,tension=0.4}{v1,v2,v1}
\fmf{photon}{r1,v1}
\fmffreeze
\fmfIR{v2}
\end{fmfgraph*}}
\eea
using a notation where the subscript zero refers to the mass of the
scalars in the loops being set to $m^2 = 0$. As mentioned at the start
of this appendix, for the purpose of regularizing this calculation
genuinely two-loop topologies can be computed in the massless
limit. However, recursively one-loop diagrams (labeled with the
superscript IR2) will exhibit IR divergences without the inclusion of
a regulator mass $\omega$. We have (retaining the superscript IR to
indicate that the full expression involves the specifically IR
regulated diagrams)
\be
\Pi^{(2,\ep)}_{0;\mu\nu} = g^4 \left (
\frac{p^{4\ep}}{\mu^{4\epsilon}}\right ) \left (
\pi^{(\UV2,\ep)}_{\YM;\mu\nu} + \pi^{(\IR2,\ep)}_{\YM;0;\mu\nu} +
\pi^{(\UV2,\ep)}_{0;\mu\nu} + \pi^{(\IR2,\ep)}_{0;\mu\nu} \right ) 
\ee
noting that it should not be interpreted that this expression is
transverse. The IR regulated gluon and scalar propagators are defined
as
\bea
G_{\mu\nu}^{(\IR1,\ep)}(q) &=& g^2 \frac{\pi^{(1,\ep)}_{\YM} +
  \pi^{(1,\ep)}_{0}}{\mu^{2\ep}(q^2 +
  \omega^2)^{\frac{3}{2}-\ep}}\left ( g^{\mu\nu} - \frac{q_\mu
  q_\nu}{q^2 + \omega^2}\right )\\
D^{(\IR1,\ep)}(q) &=& g^2 \frac{\pi^{(1,\ep)}_{\phi;0}
}{\mu^{2\ep}(q^2 + \omega^2)^{\frac{3}{2}-\ep}} .
\eea
Now, regarding the notation: at this point there are two quantities
which can be regarded as masses, $m^2$ and $\omega^2$. $m^2$ refers to
the Higgs mass which enters the problem via the scalar propagator, which
we have already set to zero. Whereas, $\omega^2$ is an unphysical
regulator mass introduced to regulate IR divergences in some two-loop
diagrams. So, for instance, the diagrams which comprise
$\pi^{(\IR2,\ep)}_{\YM;0;\mu\nu}$ are calculated using finite
$\omega^2$, but setting $m^2 = 0$. One may ask why we do not simply
regulate the IR divergences by keeping the scalar field massive from the
onset? There are two reasons.  First, a number of divergences arise from
a $1/p^3$ gauge field propagator, so this would not solve the problem
entirely.  Second, in general these diagrams are introduced to regularize
the UV divergences in the problem. To compute the leading order UV
behavior, it is sufficient to set $m^2=0$, which drastically simplifies
the majority of the diagrams which must be calculated. Then, the IR
divergences which would arise in the bare perturbation theory are
handled with $\omega^2$, of which the final results will be independent
regardless.

Defining $\chi = p/m$, the individual components are
{\footnotesize\bea
 \nonumber &&\pi^{(\UV2,\ep)}_{\YM;\mu\nu}=\frac{C_A^2}{16\pi^2}
\Bigg[ \frac{(\xi + 2)(\xi^2 + 2\xi +
11)}{48\epsilon}g_{\mu\nu} \\
\nonumber &&\quad-~ \frac{8(7\xi^3 + 75\xi^2 + 221\xi + 233) +
18\zeta(2)(\xi^2+3)(\xi^2 + 2\xi^2 + 17)}{768}
\Tproj_{\mu\nu} \\
&&\quad-~\frac{7\xi^3 + 32\xi^2 + 79\xi + 42}{48}\Lproj_{\mu\nu} \Bigg ] \label{eq:ApppitwoUVYM}\\
 \nonumber &&\pi^{(\IR2,\ep)}_{\YM;0;\mu\nu}=\frac{C_A}{16\pi^2}
\Bigg[ -\frac{4(\xi + 2) \left (\pi^{(1,0)}_{\YM} + \pi^{(1,0)}_{\HIG;0} \right )}{3\ep}g_{\mu\nu} 
+ \bigg [\frac{4\left (\pi^{(1,0)}_{\YM} + \pi^{(1,0)}_{0} \right )}{3}\bigg ( 2(\xi + 2) \log 4\chi^2 
\\
\nonumber &&\quad-~ \frac{ 
8(\xi + 2)\chi^6 + (20\xi+42)\chi^4 + 3(5\xi + 11)\chi^2 + 4(\xi + 2) }{\chi^3(\chi^2+1)^{\frac{3}{2}}} \text{arcsinh} ( \chi) \\
\nonumber &&\quad +~
 \frac{ (5\xi + 16) \chi^4 + 5 (2 \xi + 5)\chi^2 + 4 (\xi + 2)  }{\chi^2(1+\chi^2)} \bigg )+ \frac{(\xi + 2) (C_A \left (\xi^2 + 6 \xi - 6 \right) -2 T_R)}{24}  \bigg]
\Tproj_{\mu\nu}\\
\nonumber &&\quad +~ \bigg [\frac{4\left (\pi^{(1,0)}_{\YM} + \pi^{(1,0)}_{0} \right )}{3}\bigg ( 2(\xi + 2) \log 4\chi^2 - \frac{ 
4(\xi + 2)\chi^4 + 2(\xi - 1)\chi^2 - 8(\xi + 2) }{\chi^3(\chi^2+1)^{\frac{1}{2}}} \text{arcsinh} ( \chi) \\
\label{eq:ApppitwoIRYM} &&\quad +~
 \frac{(5 \xi + 6)\chi^2 - 8 (\xi + 2)  }{\chi^2} \bigg )+ \frac{(\xi + 2) \left (C_A \left (\xi^2 + 6 \xi - 6 \right) -2 T_R \right)}{24}  \bigg]
\Lproj_{\mu\nu}\\
 \nonumber &&\pi^{(\UV2,\ep)}_{0;\mu\nu}= \frac{T_R}{16\pi^2}
\Bigg[ \frac{4C_R - (\xi + 2)C_A}{12\epsilon}g_{\mu\nu} \\
\nonumber &&\quad+~ \frac{16\left (18\zeta(2) - 8(\xi - 3) \right)C_R + \left(18\zeta(2)(\xi^2 -5)+ 80\xi + 272 \right)C_A}{96}
\Tproj_{\mu\nu} \\
 &&\quad-~ \frac{2(\xi+5)C_R - (3\xi +4)C_A}{6}
\Lproj_{\mu\nu}\\
 \nonumber &&\pi^{(\IR2,\ep)}_{0;\mu\nu} = \frac{T_R}{16\pi^2}
\Bigg[ -\frac{4 \pi^{(1,0)}_{\HIG;0}}{3\ep}g_{\mu\nu} 
+ \bigg [\frac{4\pi^{(1,0)}_{\HIG;0}}{3}\bigg (2\log 4\chi^2 
- \frac{16(\chi^2+1)^{\frac{3}{2}}}{\chi^3} \text{arcsinh} ( \chi) \\
\nonumber &&\quad +~
 \frac{22\chi^2 + 16}{\chi^2} \bigg ) + \frac{(\xi -1) C_R}{2}  \bigg]
\Tproj_{\mu\nu}
+ \bigg [\frac{4\pi^{(1,0)}_{\phi;0}}{3}\bigg (2\log 4\chi^2 
- \frac{4(\chi^4 - 4 \chi^2 - 8)}{\chi^3 (\chi^2 + 1)^{\frac{1}{2}}} \text{arcsinh} ( \chi) \\
&&\quad +~
 \frac{6\chi^2 - 32}{\chi^2} \bigg ) + \frac{ (\xi -1) C_R}{2}  \bigg]
\Lproj_{\mu\nu} .
\eea}

\subsection{Two-loop Higgs self-energy}

The two-loop Higgs self-energy is specified by the diagrams
\bea
\nonumber \pi^{(\UV2,\epsilon)}_{\phi;0} &\propto&  
~
\parbox{20mm}{\begin{fmfgraph}(20,20)
\fmfleft{l1}\fmfright{r1}\fmfforce{0.25w,0.5h}{v1}\fmfforce{0.75w,0.5h}{v2}
\fmfforce{0.5w,0.75h}{vt}\fmfforce{0.5w,0.25h}{vb}
\fmf{plain}{l1,v1}
\fmf{photon,right=0.5}{v1,vb}
\fmf{photon,left=0.5}{vt,v2}
\fmf{plain,left=0.5}{v1,vt}
\fmf{plain,right=0.5}{vb,v2}
\fmf{plain}{v2,r1}\fmf{fermion}{vt,vb}
\end{fmfgraph}}
~ + ~
\parbox{20mm}{\begin{fmfgraph}(20,20)
\fmfleft{l1}\fmfright{r1}\fmfforce{0.25w,0.5h}{v1}\fmfforce{0.75w,0.5h}{v2}
\fmfforce{0.5w,0.75h}{vt}\fmfforce{0.5w,0.25h}{vb}
\fmf{plain}{l1,v1}
\fmf{photon,left=0.5}{v1,vt,v2}\fmf{photon}{vt,vb}
\fmf{fermion,right=0.5}{v1,vb}
\fmf{plain,right=0.5}{vb,v2}
\fmf{plain}{v2,r1}
\end{fmfgraph}}
~ + 2 ~
\parbox{20mm}{\begin{fmfgraph}(20,20)
\fmfleft{l1}\fmfright{r1}\fmfforce{0.25w,0.5h}{v1}\fmfforce{0.75w,0.5h}{v2}
\fmfforce{0.5w,0.75h}{vt}\fmfforce{0.5w,0.25h}{vb}
\fmf{plain}{l1,v1}
\fmf{photon,right=0.5}{vt,v2}
\fmf{fermion,left=0.5}{v1,vt}
\fmf{plain,left=0.5}{vt,v2}
\fmf{photon,left=0.5}{v2,vb,v1}
\fmf{plain}{v2,r1}
\end{fmfgraph}}\\
&& +~ 2 ~
\parbox{20mm}{\begin{fmfgraph}(20,20)
\fmfleft{l1}\fmfright{r1}\fmfforce{0.25w,0.5h}{v1}\fmfforce{0.75w,0.5h}{v2}
\fmf{plain}{l1,v1}\fmf{fermion,left=1}{v1,v2,v1}
\fmf{plain}{v1,v2}
\fmf{plain}{v2,r1}
\end{fmfgraph}}
~ + \frac{1}{2} ~
\parbox{20mm}{\begin{fmfgraph}(20,20)
\fmfleft{l1}\fmfright{r1}\fmfforce{0.25w,0.5h}{v1}\fmfforce{0.75w,0.5h}{v2}
\fmf{plain}{l1,v1}\fmf{photon,left=1}{v1,v2,v1}
\fmf{fermion}{v1,v2}
\fmf{plain}{v2,r1}
\end{fmfgraph}}
~ + ~
\parbox{20mm}{\begin{fmfgraph}(20,20)
\fmfleft{l1}\fmfright{r1}\fmfforce{0.25w,0.5h}{v1}\fmfforce{0.75w,0.5h}{v3}
\fmfforce{0.5w,0.5h}{v2}
\fmf{plain}{l1,v1}
\fmf{fermion,left=1}{v1,v2}
\fmf{photon,left=1}{v2,v1}
\fmf{plain}{v3,r1}
\fmf{fermion,left=1}{v2,v3}
\fmf{photon,left=1}{v3,v2}
\end{fmfgraph}}~~\\
 \pi^{(\IR2,\epsilon)}_{\phi;0} &\propto& 
\parbox{20mm}{\begin{fmfgraph*}(20,20)
\fmfleft{l1}\fmfright{r1}
\fmfforce{0.25w,0.5h}{v1}\fmfforce{0.75w,0.5h}{v3}
\fmfforce{0.5w,0.25h}{v2}\fmfforce{0.5w,0.75h}{v4}
\fmf{plain}{l1,v1}\fmf{plain}{v3,r1}
\fmf{photon,left=0.5,tension=0.4}{v1,v4,v3}
\fmf{fermion,right=1}{v1,v3}
\fmfIR{v4}
\end{fmfgraph*}}
~ + ~
\parbox{20mm}{\begin{fmfgraph*}(20,20)
\fmfleft{l1}\fmfright{r1}
\fmfforce{0.25w,0.5h}{v1}\fmfforce{0.75w,0.5h}{v3}
\fmfforce{0.5w,0.25h}{v2}\fmfforce{0.5w,0.75h}{v4}
\fmf{plain}{l1,v1}\fmf{plain}{v3,r1}
\fmf{photon,left=1}{v1,v3}
\fmf{fermion,right=0.5}{v1,v2}
\fmf{plain,right=0.5}{v2,v3}
\fmfIR{v2}
\end{fmfgraph*}}
~+ \frac{1}{2}~
\parbox{20mm}{\begin{fmfgraph*}(20,20)
\fmfleft{l1}\fmfright{r1}\fmftop{t1}\fmfforce{0.5w,0.5h}{v1}
\fmfforce{0.5w,0.5h}{v3}
\fmf{plain}{l1,v1}\fmf{plain}{v1,r1}
\fmf{phantom,tension=5.0}{t1,v2}
\fmf{photon,left=1,tension=0.4}{v1,v2,v1}
\fmffreeze
\fmfIR{v2}
\end{fmfgraph*}}
~ + 2 ~
\parbox{20mm}{\begin{fmfgraph*}(20,20)
\fmfleft{l1}\fmfright{r1}\fmftop{t1}\fmfforce{0.5w,0.5h}{v1}
\fmfforce{0.5w,0.5h}{v3}
\fmf{plain}{l1,v1}\fmf{plain}{v1,r1}
\fmf{phantom,tension=5.0}{t1,v2}
\fmf{fermion,left=1,tension=0.4}{v1,v2,v1}
\fmffreeze
\fmfIR{v2}
\end{fmfgraph*}}
\eea
where once again IR divergences are handled with a regulator mass
$\omega$. Including a counter-term, we have
\be
\Pi^{(2,\ep)}_{\phi;0} = g^4 \left ( \frac{p^{4\ep}}{\mu^{4\epsilon}}\right ) \left ( \pi^{(\UV2,\ep)}_{\phi;0} + \pi^{(\IR2,\ep)}_{\phi;0} \right ) - \delta m^2
\ee
with
{\footnotesize\bea
 \nonumber &&\pi^{(\UV2,\ep)}_{\phi;0}=\frac{1}{16\pi^2}
\Bigg[ \frac{C_R \big (C_A(\xi-1)(\xi+3) + 4C_R(2\xi-3) \big) - 16 (1 + d_R)x^2}{16\ep}\\
\nonumber &&\quad+~\frac{C_R \Big [ C_A \big (12\zeta(2) + 3\xi^2 + 22\xi + 27 \big)- 4C_R \big (18\zeta(2) + \xi^2 + 6\xi - 1 \big ) \Big ]}{16} \\
&&\quad + ~ 6(1+d_R)x^2\Bigg ]\\
 \nonumber &&\pi^{(\IR2,\ep)}_{\phi;0} = 
\frac{(1+d_R)x}{8\pi^2} \Bigg [ \frac{2\pi^{(1,0)}_{\HIG;0}}{\ep} - 4\pi^{(1,0)}_{\HIG;0} \left (\log 4\chi^2 - 1\right) + \frac{(1-\xi)C_R}{2} \Bigg ]
\\
\nonumber&& \quad+~ \frac{C_R}{16\pi^2}\Bigg[\frac{4 \left ( \pi^{(1,0)}_{\YM} + \pi^{(1,0)}_{0} \right ) - 2\xi \pi^{(1,0)}_{\phi;0}}{\ep} 
- \bigg [ 8 \left ( \pi^{(1,0)}_{\YM} + \pi^{(1,0)}_{0} \right ) - 4 \xi \pi^{(1,0)}_{\HIG;0} \bigg ]\log 4\chi^2 
\\
\nonumber &&\quad+ ~ \frac{ 8 \bigg [ \left ( \pi^{(1,0)}_{\YM} + \pi^{(1,0)}_{0} \right )\left (8\chi^4 + 15 \chi^2 + 6\right) + 3\pi^{(1,0)}_{\HIG;0} (\chi^2 + 1)\left ( 3\xi (\chi^2 + 1) - 1\right ) \bigg]}{3\chi(\chi^2+1)^{\frac{3}{2}}} \text{arcsinh} ( \chi) \\
\nonumber &&\quad -~ \frac{1}{24}\bigg[32
 \left ( \pi^{(1,0)}_{\YM} + \pi^{(1,0)}_{0} \right ) \frac{\chi^2 + 3}{\chi^2 + 1} + 96 \pi^{(1,0)}_{\HIG;0}(7\xi - 2) \\
&&\quad+~ 3C_A (\xi^2 + 6\xi - 6) - 6T_R - 12C_R \xi(\xi - 1)
\bigg]\\
 && \delta m^2 = \frac{1}{16\pi^2\epsilon} \bigg [\frac{C_R(7 C_A - 6C_R - 2T_R)}{8}g^4 + C_R(d_R + 1)g^2\lambda - (d_R + 1)\lambda^2 \bigg ]
\eea}
\end{fmffile}
Due to the counter-term, the scale dependence of $m^2$ is given by the
RG equation
\be
\frac{d m^2}{d \log \mu} = \beta_{m^2}(g^2,\lambda)
\ee
with
\be
\beta_{m^2}(g^2,\lambda) = - \frac{\partial \delta m^2}{\partial
  g^2}\frac{d g^2}{d\log\mu} - \frac{\partial \delta m^2}{\partial
  \lambda}\frac{d \lambda}{d\log\mu} = -2\epsilon g^2 \frac{\partial
  \delta m^2}{\partial g^2} - 2\epsilon \lambda \frac{\partial \delta
  m^2}{\partial \lambda} .
\ee
For instance, with an $\text{SU}(2)$ fundamental Higgs in Landau gauge,
\be
\beta_{m^2}(g^2,\lambda) = - \frac{1}{16\pi^2}\bigg [ \frac{51}{16}g^4
  + 9g^2 \lambda - 12\lambda^2\bigg ] .
\ee

\bibliography{SUNHiggsBib}
\bibliographystyle{JHEP}

\end{document}